\documentclass[twocolumn]{aastex61}

\received{TBD}
\revised{TBD}
\accepted{TBD}
\submitjournal{AAS Journals}

\shorttitle{$K_s$-band photometry of RR~Lyrae Stars}
\shortauthors{Layden \textit{et al.}}
\begin{document}

\title{Infrared $K_s$-band Photometry of Field RR~Lyrae Variable Stars}

\correspondingauthor{Andrew Layden}
\email{laydena@bgsu.edu}

\author[0000-0002-6345-5171]{Andrew C. Layden}
\affiliation{Bowling Green State University\\
%104 Overman Hall\\
Physics \& Astronomy Department\\
Bowling Green, OH 43403, USA}

\author{Glenn P. Tiede}
\affil{Bowling Green State University\\
%104 Overman Hall\\
Physics \& Astronomy Department\\
Bowling Green, OH 43403, USA}

\author[0000-0003-3096-4161]{Brian Chaboyer}
\affil{Dartmouth College\\
Department of Physics and Astronomy\\
Hanover, NH 03784, USA}

\author{Curtis Bunner}
\affil{Bowling Green State University\\
%104 Overman Hall\\
Physics \& Astronomy Department\\
Bowling Green, OH 43403, USA}
%\affil{Institution\
%Columbus, OH 43403, USA}

\author{Michael T. Smitka}
\affil{The Pennsylvania State University\\
%104 Davey Lab\\
Department of Physics\\
University Park, PA 16802, USA}

%\author{Bruce W. Carney}
%\affil{University of North Carolina\\
%}

% 
% 
% 

%\collaboration{(AAS Journals Data Scientists collaboration)}

%\affiliation{AAS Journals Associate Editor-in-Chief}
%\nocollaboration

%\author{Folks from PROMPT}
%\altaffiliation{Creator of AASTeX v6.1}
%\collaboration{(LaTeX collaboration)}



\begin{abstract}

%[TO DO: continue to develop this as the conclusions solidify.]
%\textbf {Abstract is largely new... }
We present multi-epoch infrared photometry in the $K_s$-band for  74 bright RR~Lyrae variable
stars tied directly to the 2MASS photometric system.  We systematize additional $K$-band photometry from the literature to the 2MASS system and combine it to obtain photometry for 146~RR~Lyrae stars on a consistent, modern system. A set of outlier stars in the literature photometry is identified and discussed.  Reddening estimates for each star were gathered from the literature and combined to provide an estimate of the interstellar absorption affecting each star, and we find excellent agreement with another source in the literature.  We utilize trigonometric parallaxes from the Second Data Release of ESA's \textit{Gaia} astrometric satellite to determine the absolute magnitude, $M_{K_s}$ for each of these stars, and analyze  them using the astrometry based luminosity prescription to obtain a parallax-based calibration of $M_K$(RR).   Our period-luminosity-metallicity relationship is 
$ M_{K_s} = (-2.8\pm 0.2)  (\log P +0.27) + (0.12\pm 0.02) ( [\mathrm{Fe/H}] + 1.3) - (0.41\pm 0.03)$ mag.  
%$\mathbf{ M_{K_s} = (-2.8\pm 0.2)  (\log P +0.27) + (0.12\pm 0.02) ( [\mathrm{Fe/H}] + 1.3) - (0.41\pm 0.03)}$ .  
A \textit{Gaia} global zero-point error of $\pi_{zp} = -0.042\pm 0.013\,$mas   is determined for this sample of RR~Lyrae stars. 

% Original manuscript: Our best period-luminosity-metallicity relationship is  
%$M_{K_s} = (-2.5\pm 0.3)  (\log P +0.27) + (0.04 \pm 0.02)(  [\mathrm{Fe/H}] + 1.3)  - (0.42\pm 0.04)$ mag.  We also present provisional analysis of subsets of the RRL sample in which we separate stars with halo and disk kinematics, and separate the halo stars into subsets based on their period-shift which we identify with the Oosterhoff~I and II populations seen in globular clusters.  We also find that a relationship between period, luminosity, and period-shift is an equally valid way of describing the $M_K$(RR) relation.

%We conclude with recommendations for how additional $K_s$ magnitudes might be obtained to maximize the effective use of parallaxes from future Data Releases from the \textit{Gaia} satellite.

%TBD.  The AAS Journals, the Astrophysical Journal (ApJ), the Astrophysical Journal Letters (ApJL), and Astronomical Journal (AJ), all have a 250 word limit for the abstract.  If you exceed this length the Editorial office will ask you to shorten it. [this %bit is ~40 words]

\end{abstract}

% Other AAS keyword options for this paper:  methods: observational --- stars: distances --- stars: horizontal branch --- stars: luminosity function, mass function --- stars: Population II --- stars: statistics --- 
%ADS \keywords{methods: data analysis --- stars: individual (KIC~8462852) --- stars: peculiar --- stars: variable: general }



% remember  $M_{\sun}$     $\gtrsim$0.15   $\lesssim$0.02  

% ToDo: final search/replace cleanup of temporary acronyms:  RRL --> RR~Lyrae (stars)

\section{Introduction} \label{sec_intro}

For many decades, the RR Lyrae variable stars (RRL) have  % been the cornerstone of (
played a fundamental role in establishing the distance scale for old stellar populations, including globular clusters, the Galactic bulge, and the Magellanic clouds (e.g., see the review by \citet{cacciari12}).  The resulting calibrations of the visual absolute magnitude, $M_V$(RR) and its near-infrared analog, $M_K$(RR), provide a critical lower rung on the intergalactic distance ladder \citep{dambis13, clementini17}.  
These stars have also proved useful in mapping 
%three-dimensional extended objects including 
Galactic stellar overdensities and streams, and interacting galaxies (e.g., \citet{vivas06, bmc}).
Furthermore, accurate distance measurements to globular clusters and nearby galaxies are critical to determining the main-sequence turn-off ages to these systems, with a 1\% uncertainty in distance leading to $\sim$2\% uncertainty in the derived age of a given system \citep{chaboyer96}.  Accurate ages for globular clusters can set a stringent limit on the age of the Universe, providing an important independent check on that age from recent precision cosmology studies \citep{planck18}, and they can be used to look for trends in the chronology and metal-enrichment histories of the old stellar populations in our Galaxy, thereby constraining models of its formation and early evolution \citep{dekany18}. 

% LMC distance 2009 \citet{borissova})

Previous calibrations of the RRL absolute magnitude have taken a variety of approaches, including utilizing the Baade-Wesslink and infrared flux methods (e.g., \citet{liu_janes90}, \citet{jones92}, \citet{skillen93}, and references therein), statistical parallax (e.g, \citet{layden96, fernley98, dambis13} and references therein), and fitting the main sequences of globular clusters to field subdwarfs with high-quality parallaxes \citep{carretta00}.
%or to theoretical isochrones \citep{wagner18}.
%a host of other methods, some of which are discussed in \citet{cacciari12}. 
Despite these efforts, a high degree of uncertainty remains.  For example, the $M_K$(RR) calibrations listed in Table~2 of \citet{cacciari12} range over 0.14 mag at a fiducial metallicity.

Direct trigonometric parallax is the preferred method of obtaining the distance to any stellar source, yet even using precision astrometry from the \textit{Hipparcos} satellite and the \textit{Hubble Space Telescope}, the few RRL closest to the Sun have not yet provided a definitive RRL luminosity calibration (e.g., see Table~1 of \citet{clementini17} and references therein).  
%\textbf{The rest of the Intro is new...}
%remained tantalizing outside the reach of direct parallax.  
It was with these ideas in mind that our team proposed to use NASA's \textit{Space Interferometry Mission PlanetQuest (SIM-PQ)} astrometric satellite \citep{unwin08}, which was designed to deliver parallaxes with precisions at the micro-arcsecond level, in order to determine the distance scale to Population~II objects. Our key project was awarded 1330 hours of observing time on \textit{SIM-PQ} to obtain parallaxes for 21 globular clusters, 60 field RRL, and 60 metal-poor field subdwarf stars \citep{chaboyer05}.  A goal of the project was to calibrate the RRL absolute magnitude scales, $M_V$(RR) and $M_K$(RR), with unprecedented precision.  At that time, we developed a list of potential target RRL, and began ground-based photometric time-series observations in the $VI_CK_s$ passbands to provide phased light curves and mean apparent magnitudes in support of the space-based astrometry.  Unfortunately, delays in the \textit{SIM-PQ} program led to cost overruns, and though a simplified version of the mission called \textit{SIM-Lite} \citep{marr10} briefly replaced it, the project was eventually discontinued at the end of 2010.   This paper will focus on our $K_s$ observations and the resulting $M_{K_s}(RR)$ calibration, along with comparisons with infrared work by other researchers. 
% See https://science.nasa.gov/missions/sim

Though many studies have pushed forward with the goal of refining the 
%$M_V$(RR) and $M_K$(RR) 
RRL luminosity calibration in the intervening years, a revolutionary advancement through space-based precision astrometry has been absent until recently.  Fortunately, this situation is changing rapidly as results from the European Space Agency's \textit{Gaia} satellite \citep{gaia} become available.  Preliminary parallaxes from the \textit{Gaia} Data Release~1 \citep{gaiaDR1} for several hundred field RRL were analyzed by \citet{clementini17}, who used single-epoch $K_s$-band apparent magnitudes and interstellar reddening values compiled by \citet{dambis13} %for several hundred RRL 
to determine preliminary infrared period-luminosity (PL) and period-luminosity-metallicity (PLZ) relations.  The second data release, DR2, 
%from \textit{Gaia} 
\citep{gaiaDR2cat} contained significantly improved astrometry, and was used by \citet{muraveva18}, again using photometric data from \citet{dambis13}, to further improve the infrared PL and PLZ relations.  Still further improvements in the \textit{Gaia} parallaxes are expected in the third and final data releases, which are currently scheduled for 2020 and 2022, respectively.

% Clementinin17 DR1 paper uses RRL compilation including ZW-metallicity and AV for 402 stars by Dambis, A. K., Berdnikov, L. N., Kniazev, A. Y. et al. 2013, MNRAS, 435, 3206
% Gaia: mission overview  \bibitem[\textit{Gaia} Collaboration et al.(2016a)] {gaia} \textit{Gaia} Collaboration, Prusti, I., de Bruijne, J. H. J., et al. 2016a, \aap, A1
% DR1: summary of the astrometric, photometric, and survey properties \bibitem[\textit{Gaia} Collaboration et al.(2016b)] {gaiaDR1} \textit{Gaia} Collaboration, Brown, A. G. A., Vallenari, A., et al. 2016b, \aap, A2
% DR2: summary of the contents and survey properties \bibitem[\textit{Gaia} Collaboration et al.(2018)] {gaiaDR2} \textit{Gaia} Collaboration, Brown, A. G. A., Vallenari, A., et al. 2018, arXiv: 1804.09365
% DR2: validation of Cepheid and RRL \bibitem[Clementini et al.(2018)] {clementini18} Clementini, G., Repepi, V., Molinaro, A., et al. 2018, arXiv:1805.02079
% RRL as standard candles in Gaia DR2  \bibitem[Muraveva et al.(2018)]{muraveva18}Muraveva, T., Delgado, H.E., Clementini, G., Sarro, L.M. \& Garofalo, A.\ arXiv:1805.08742

% Also check out Dambis, A. K., Rastorguev, A. S., \& Zabolotskikh, M. V. 2014, MNRAS, 439, 3765 does RRL abs-mag calibration in WISE W1 passband

As these improvements unfold, we felt it was timely to contribute our multi-epoch apparent $K_s$ photometry of field RRL and integrate them with the existing photometry for these stars to provide an optimal photometric database for present and future PL and PLZ analyses.  In Section~\ref{sec_sel} of this paper, we describe the selection of our sample of RRL and compile pre-existing data including interstellar reddening estimates, pulsation periods, and metallicities.  In Section~\ref{sec_obs} we describe our infrared imaging observations, while in Section~\ref{sec_calib} we present our methods for photometric measurement and calibration.  Our method for fitting the phased light curves with templates of characteristic shape is presented in Section~\ref{sec_tmplt}, and details concerning the photometric uncertainties are assessed in Section~\ref{sec_errs}.  In Section~\ref{sec_compare} we compare our photometry with that of other sources and integrate them into a comprehensive database useful for future studies.  We utilize the DR2 parallaxes in Section~\ref{sec_gaia} using advanced statistical methods to determine our version of the PLZ relation.
%, and explore several ways of subdividing the RRL sample to assess whether different RRL subpopulations have measureably different luminosities.  
We present our conclusions in Section~\ref{sec_concl}, and in an Appendix we provide 
%details about our analysis of interstellar reddening, our separation of sub-populations, and our 
notes on the light curves of individual stars.

\section{Sample Selection}  \label{sec_sel}

Because \textit{SIM-PQ} was designed as a targeted instrument rather than an all-sky survey satellite like \textit{Gaia}, we developed a target list for our $K_s$-band photometry program based on the 144 RRL in \citet{fernley98} and supplemented it with stars from the RRL lists of \citet{layden94} and other sources, resulting in a working database of 172 RRL.  Careful prioritization based on astrophysical properties like metallicity, period, Oosterhoff group, and kinematics allowed us to focus attention on the most relevant stars.  We also prioritized stars with no $K$-band photometry in the literature, and stars whose  literaturephotometry appeared to be of lower quality. We tended to avoid stars that were known to exhibit the Blazhko effect.  Due to a variety of factors including sky visibility and weather, we ultimately obtained $K_s$-band photometry for the first 75 stars listed in Table~\ref{tab_targets}. 
%lists the stars we selected for our {\it SIM} project.  
The column labeled ``ID (2MASS)'' contains each star's unique identifier in the Two-Micron All-Sky Survey (2MASS) of \citet{skrutskie06}, which is the sexigesimal equatorial coordinates for the star coded in the standard catalog format.  The next columns contain each star's Galactic longitude and latitude ($l$, $b$) %Galactic longitude, $l$, and latitude, $b$,  
in degrees.
%, from which we determined the star's interstellar reddening, $E(B-V)$, using the NEW (download it!) revised dust map of Schlegel et al REF.  
For most stars, we obtained two estimates of the period, one from 
\citet{fernley98} and a second from a recent inspection of the International Variable Star Index (VSX)\footnote{Accessed circa 2018 January from \url{https://www.aavso.org/vsx/index.php}.}.  
 These two values are usually very similar, and are reported in 
%columns five and six of 
Table~\ref{tab_targets} as $P_{F98}$ and $P_{VSX}$, respectively.

% TABLE 1
\begin{deluxetable}{lcrrllccccccccl}
%\tabletypesize{\small}
\tabletypesize{\scriptsize}
%\begin{deluxetable}{cccrrrrrr}
%aastex52\tabletypesize{\scriptsize}
%aastex52 
\rotate
\tablecaption{The RR~Lyrae Sample and Literature Data \label{tab_targets}}
% table1_sub2.txt --> tab1b.txt at submission
%aastex52 \tablewidth{0pt}
\tablehead{
\colhead{Star} & \colhead{ID (2MASS)} & \colhead{$l$} & \colhead{$b$}  & 
\colhead{$P_{F98}$} & \colhead{$P_{VSX}$} &
\colhead{[Fe/H]$_{L94}$} & \colhead{[Fe/H]$_{F98}$}  & \colhead{$N_{src}$} & 
\colhead{$E_{SF11}$}  &  \colhead{$E_{B92}$}  &\colhead{$E_{F98}$}  &
\colhead{$K_{F98}$} & \colhead{Ref\tablenotemark{a}} 
& \colhead{Comment\tablenotemark{b} }
}
%\colnumbers
\startdata
DM~And  & 23320072+3511488 & 105.11 & --24.91 & 0.630389 &  0.6304244 & --2.32 &  \nodata  & 0 & 0.081 & \nodata & \nodata & \nodata & \nodata & \nodata \\
WY~Ant   & 10160494--2943423 &  266.94 & +22.08 & 0.574312  & 0.5743427 & --1.66 & --1.48 & 2 & 0.059 & 0.05 & 0.07 & 9.64 & 1 & bw36 \\
%RU~Bov & 12:34:56.78 & 09:87:65.4 & 12.34 & 0.28 & 123.45 & 87.65 & 0.123 & 0.567890 & --1.23 & 1,3,5 \\
\nodata & \nodata & \nodata & \nodata & \nodata & \nodata & \nodata & \nodata & \nodata & \nodata & \nodata & \nodata & \nodata & \nodata  & \nodata  \\
AU~Vir   & 13244801--0658455 & 317.45 & +54.95 & 0.343230 & 0.3432307 & \nodata & --1.50 & 3 & 0.025  & \nodata & --0.01 & 10.81 & 3 & RRc  \\
AV~Vir  & 13201156+0911163 & 325.01 & +70.82 & 0.656908 & 0.6569073 & --1.32 & --1.25 & 2 & 0.027 & \nodata & --0.01  & 10.63 & 2 & \nodata \\
%NSV~660 & 01545017+0015007 & 154.61 & --58.67 & \nodata & 0.636589 & \nodata &  --1.31 & 0 & 0.031 & \nodata & \nodata & ??? & \nodata & Szabo  \\
NSV~660 & 01545017+0015007 & 154.61 & --58.67 & \nodata & 0.636589 & \nodata &  \nodata  & 0 & 0.031 & \nodata & \nodata & \nodata  & \nodata & \citet{szabo14}  \\
SW~And &  00234308+2924036 & 115.72 & --33.08 &  0.442279 & 0.442262 & --0.38 & --0.24 & 6 & 0.039 & 0.09 & 0.07 & 8.54 & 1 & B?  bw10, bw25 \\
\nodata & \nodata & \nodata & \nodata & \nodata & \nodata & \nodata & \nodata & \nodata & \nodata & \nodata & \nodata & \nodata & \nodata  & \nodata \\
AF~Vir &   14280981+0632438 & 355.47 & +59.16 & 0.483722 & 0.483747 & --1.46 & --1.33 & 2 & 0.021 & \nodata & 0.05 & 10.75 & 3 & \nodata \\
BN~Vul  & 19275606+2420505 &    58.63  & +3.41 & 0.594125 & 0.5941318 & \nodata & --1.61 & 1 & 1.053 & 0.44  & 0.41 & 8.81 & 2 & \nodata \\   
\enddata
%\tablenotetext{a}{Reddening references: 1 = Jones, 2 = Smith, etc.}
%\tablenotetext{a}{Period references: 1---VSX, 2---\citet{fernley98}, 3---This work, 4---\citet{szabo14}. }
%\tablenotetext{b}{Metallicity references: 1---\citet{fernley98}, 2---\citet{layden94}, 3---\citet{szabo14}. }
\tablenotetext{a}{Photometry references: 1---\citet{skillen93}, 2---\citet{fernley93b}, 3---Unpublished (see \citet{fernley98}).  A reference to the source of the original photometry for stars with Baade-Wesselink analyses are given in the Comments column, where \textbf{bw10} = \citet{liujanes89},   \textbf{bw21} = \citet{jones87a}, 
\textbf{bw22} = \citet{jones87b}, \textbf{bw23} = \citet{jones88a}, \textbf{bw24} = \citet{jones88b}, \textbf{bw25} = \citet{jones92}, 
\textbf{bw31} = \citet{fernley89x}, \textbf{bw32} = \citet{skillen89}, \textbf{bw33} = \citet{fernley90a}, \textbf{bw34} = \citet{fernley90b}, \textbf{bw35} = \citet{cacciari92}, and \textbf{bw36} = \citet{skillen93}.  }
%, and \textbf{Szabo} = \citet{szabo14}.  }
%with Baade-Wesselink analyses are given in the Comments column, where ``bw10'' = \citet{liujanes89},   ``bw21'' = \citet{jones87a}, 
%``bw22'' = \citet{jones87b}, ``bw23'' = \citet{jones88a}, ``bw24'' = \citet{jones88b}, ``bw25'' = \citet{jones92}, 
%``bw31'' = \citet{fernley89x}, ``bw32'' = \citet{skillen89}, ``bw33'' = \citet{fernley90a}, ``bw34'' = \citet{fernley90b}, ``bw35'' = \citet{cacciari92}, ``bw36'' = \citet{skillen93}, and ``Szabo'' = \citet{szabo14}.  }
%4---\citet{monson17} recalibration of original photometry from \citet{fernley89} [X~Ari], \citet{skillen89} [DX~Del], \citet{fernley90a} [V445~Oph, SS~Leo \& VY~Ser], \citet{fernley90b} [DH~Peg],. }
\tablenotetext{b}{Comment: RRc---First-ovetone, c-type pulsator; B---Exhibits the Blazhko effect in optical light; ``B?'' indicates the Blazhko effect is suspected.  For the star NSV~660 (LP~Cet), multi-epoch photometry in the 2MASS $K_s$ band is available from \citet{szabo14}, along with values for period and [Fe/H]. }
% ``OoI'' or ``OoII"---Oosterhoff group I or II respectively.} 
\tablecomments{(This table is available in its entirety in machine-readable form.)}
%\tablecomments{Table \ref{tab_targets} is published in its entirety in the electronic edition of the \textit{Astronomical Journal}. A portion is shown here for guidance regarding its form and content.}
\end{deluxetable}

% [X~Ari] \citet{fernley89} Fernley, J. A., Lynas-Gray, A. E., Skillen, I., Jameson, R. F., Marang, F., Kilkenny, D., and Longmore, A. J. 1989, \mnras, 236, 447
% [DX~Del] \citet{skillen89} Skillen, I., Fernley, J. A., Jameson, R. F., Lynas-Gray, A. E., and Longmore, A. J. 1989, \mnras, 241, 281
% [V445~Oph, SS~Leo \& VY~Ser] \citet{fernley90a} Fernley, J. A., Skillen, I., Jameson, R. F., Barnes, T. G., Kilkenny, D., and Hill, G. 1990, \mnras, 247, 287
% [DH~Peg] \citet{fernley90b} Fernley, J. A., Skillen, I., Jameson, R. F., and Longmore, A. J. 1990, \mnras, 242, 685
% [UU~Cet, RV~Phe, \& W~Tuc] \citet{cacciari92}  bib exists
% [WY~Ant, W~Crt, RV~Oct, \& BB~Pup] \citet{skillen93}  bib exists

\subsection{Stellar Abundances from the Literature} \label{ssec_samz}

 There are two principal sources of metallicities available in the literature for the RRL in Table~\ref{tab_targets}: the spectroscopic study of \citet{layden94}   (which was calibrated to the globular cluster metallicity scale of \citet{zinnwest84})  and the compilation of [Fe/H] sources by \citet{fernley98}, which includes values from \citet{layden94}.   These sources provided most of the [Fe/H] values for the bright RRL in the catalog of \citet{beers00}.  \citet{dambis13} drew their metallicities from this catalog, and the Dambis values were used in turn by the recent RRL PLZ studies utilizing \textit{Gaia} astrometry \citep{clementini17,muraveva18}.  
These values are listed in Table~\ref{tab_targets} as [Fe/H]$_{L94}$ and [Fe/H]$_{F98}$ respectively, and the column labeled $N_{src}$ lists the number of metallicity estimates from independent sources in the \citet{fernley98} compilation.   
% Moved to Sec. 2.3: \citet{szabo14} list $[\mathrm{Fe/H}] = -1.31 \pm 0.03$ dex for NSV~660 but note the possibility of systematic uncertainties, so we adopt an uncertainty of 0.10 dex.

% [Beers et al. 2000] {beers00}  Beers. T. M., Chiba, M., Yoshii, Y., Platais, I., Hanson, R. B., Fuchs, B., \& Rossi, S. 2000, \aj, 119, 2866

\begin{figure}
%aastex52 \epsscale{.80}
%\plotone{f98p_feh2b.eps}
\plotone{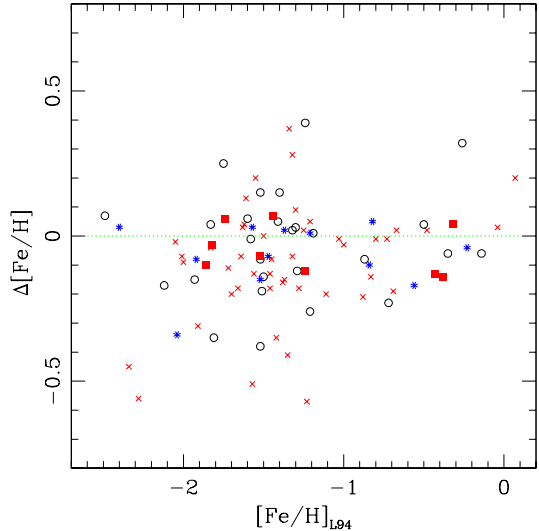}
\caption{The difference between the [Fe/H] values of \citet{layden94} and \citet{fernley98} 
%(in the sense L94 minus F98) 
is plotted as a function of each star's metallicity from \citet{layden94}.  Symbols indicate the number of literature sources utilized in the compilation of \citet{fernley98}: crosses indicate $N_{src} = 2$, circles mark $N_{src} = 3$, asterisks mark $N_{src} = 4$, and squares mark $N_{src} > 4$.  Stars having more sources scatter less around the mean. \label{fig_feh2b}}
\end{figure}

Figure~\ref{fig_feh2b} shows the relationship between these metallicities 
%of \citet{layden94} and \citet{fernley98} 
for the 93 stars in our working database of 172 objects that have both metallicity measures, where the difference $\Delta$[Fe/H] is calculated in the sense 
Layden minus Fernley.
%\citet{layden94} minus \citet{fernley98}.  
%The symbols indicate the number of studies used as sources for each star in the compilation of \citet{fernley98} -- for example, \citet{fernley98} obtained the [Fe/H] value for the star SW~And by combining the values from six separate studies.  
It is apparent that stars utilizing data from a larger number of sources ($N_{src}$) have a smaller scatter in this figure, and the statistics in Table~\ref{tab_zstat} support this claim.  Specifically, as $N_{src}$ increases from two to five-or-more sources, the standard deviation $\sigma$ of the metallicity difference decreases from 0.20 to 0.08 dex.  Because the ranking of [Fe/H] values compiled from multiple sources is demonstrably better than those of a single source, we adopt the \citet{fernley98} values as the primary source of our metallicities from Table~\ref{tab_targets}, and when absent, we use the value from \citet{layden94}.  In total, there were 21 stars for which the \citet{layden94} was the only source.
%[ACL MUST CHECK: DOES THIS ALWAYS OCCUR WHEN Nsrc=1?  No, of the Nsrc=1 cases in F98, nine occur when the sole source is L94, and 14 occur when the sole source is some other refer.  There are twelve additional cases where L94 was used b/c no value in F98, hence Nsrc=0.

The mean metallicity difference, shown in the second column of Table~\ref{tab_zstat}, is slightly negative for each line.  Depending on the weighting scheme utilized to combine them, the typical value is about $-0.06$ dex in the sense that metallicities from \citet{fernley98} are systematically more metal-rich than those of \citet{layden94}.  % HERE WE HAVE TO CHOOSE HOW TO HANDLE THE ABUNDANCES:  VERSION 1:  
This difference is small, and in past works \citep{beers00, dambis13} the values from these two sources have been utilized without any systematic corrections.  
We too are inclined to combine the values from the two sources without correction, but note that shifting the Layden values onto the Fernley system, or \textit{vice versa}, are also reasonable approaches.
%In this work, we also combine values from the two sources without correction, and in Sec.~\ref{sec_gaia} we test the impact of how these literature metallicities are combined and weighted on our PLZ relations.  We note that a careful
Clearly, a well-documented combination of \textit{all} the currently available literature values of [Fe/H] for \textit{all} field RRL is warranted.  

%VERSION 2:  Because the metallicities in \citet{layden94} are calibrated to the globular cluster scale of \citet{zinnwest84}, we have elected to subtract 0.06 dex from the [Fe/H] values of \citet{fernley98} to to shift them to match this scale.  

%VERSION 3 [THIS WAS DONE IN THE ORIGINAL SUBMISSION]:  To bring these sources into systematic agreement on the \citet{fernley98} scale, we have added 0.06 dex to the [Fe/H] values of \citet{layden94}.

%[FOR V2-3, NEED TO DECIDE HOW TO REPRESENT [Fe/H] IN TABLE 1: AS-IS OR SHIFTED TO FINAL].

The standard deviations $\sigma$ in Table~\ref{tab_zstat} contain information about the typical uncertainties in the [Fe/H] measures of both \citet{layden94} and of \citet{fernley98}.  Table~9 of \citet{layden94} reports two  estimates of the [Fe/H] uncertainty for each star.  For each group of $N_{star}$ stars from a line shown in Table~\ref{tab_zstat}, we calculated the mean of these observational uncertainties and present it as $\overline{\sigma_{L94} }$.  We then subtracted that in quadrature from the value of $\sigma$ to obtain an estimate of the typical uncertainty in the [Fe/H] value from \citet{fernley98} for stars in that group,  $\overline{\sigma_{F98} }$.  Statistically, these values may be used to weight the impact of different stars in the overall PLZ relation.  Notice that for two groups, the estimated uncertainty $\overline{\sigma_{L94} }$ accounts for all of the dispersion in the metallicity difference, so we set the entries for $\overline{\sigma_{F98} }$ to zero for these lines in Table~\ref{tab_zstat}, even though it is unlikely the Fernley metallicities have no observational uncertainty.  

%Given this evidence for increasing abundance precision with number of literature sources, we propose the following scheme for weighting the contribution of individual stars' [Fe/H] values to the PLZ fits described in Sec.~\ref{sec_gaia}.  
The data in Table~\ref{tab_zstat} are broadly consistent with the behavior $\sigma_{[\mathrm{Fe/H}]} = 0.22~N_{src}^{-0.5}$ dex for stars having metallicities in \citet{fernley98}.  For the stars with metallicities from \citet{layden94} alone we suggest $\sigma_{[\mathrm{Fe/H}]} = 0.15$ dex, based on the $\overline{\sigma_{L94} }$ values in Table~\ref{tab_zstat} and statements made in \citet{layden94}.  These values should be suitable for statistical weighting and are not intended to represent the formal uncertainty in the abundance of any given star.  
In Sec.~\ref{sec_gaia} we will test how our choice of [Fe/H] weighting, and how our choice of systematic shifts between metallicity scales, affect the PLZ relations we derive.

%[NB: BRIAN, IN SEC.8 I SUGGEST WE USE FeH ERRORS OF 0.16, 0.13, 0.10, and 0.07 dex for stars WITH Nsrc=2,3,4,>4 (respectively).  TABLE 2 MAKES IT CLEAR WHERE WE GOT THOSE VALUES FROM, ESPECIALLY IMPORTANT IF WE SHOW THE FeH ERROR BARS IN THE NEW SEC.8 FIGURES.]

% TABLE 2
\begin{deluxetable}{cccrcc}
%\begin{deluxetable}{cccrrrrrr}
%aastex52\tabletypesize{\scriptsize}
%aastex52 
%\rotate
\tablecaption{Comparing Literature Metallicities \label{tab_zstat}}
%aastex52 \tablewidth{0pt}
\tablehead{
\colhead{$N_{src}$}  & \colhead{ $\overline{\Delta \rm{ [Fe/H] } }$ } & \colhead{$\sigma$}  & \colhead{$N_{star}$}  
& $\overline{\sigma_{L94} }$  &  $\overline{\sigma_{F98} }$  }
%\colnumbers
\startdata
2      & --0.096     &    0.200 & 45   & 0.124 & 0.157 \\
3      & --0.026       &  0.181  & 27  & 0.127 & 0.129 \\
4      & --0.068     &    0.108 & 12 & 0.135 & 0 \\
$>$4 & --0.047     &    0.080  & 9 & 0.112 & 0 \\
all    &  --0.067     &    0.179 & 93 & 0.125 & 0.128 \\
\enddata
%\tablenotetext{a}{Reddening references: 1 = Jones, 2 = Smith, etc.}
%\tablecomments{(This table is available in its entirety in machine-readable form.)}
%\tablecomments{(NOTE: IF THIS IS APPROVED, COMPUTE sigmaL94 for Nsrc=1 L94-only stars.)}
\end{deluxetable}

%When the value was available for a given star, we adopted the metallicity value [Fe/H] from \citet{fernley98}, which was compiled from a large number of sources including \citet{layden94}.  When this value was not available, we adopted the [Fe/H] value from \citet{layden94} after adding 0.06 dex, a correction factor determined by regressing values common to the two sources.  This factor also serves to shift the data from the \citet{zinnwest84} metallicity scale used by \citet{layden94} to that of the \citet{gratton04} scale, as described in \citet{clementini17}.  The metallicity value on the \citet{gratton04} scale and its source reference is listed in Table~\ref{tab_targets} for each star.  The uncertainty in the metallicity of a typical star is 0.15 to 0.20 dex \citep{layden94, clementini17}.  ADD DISCUSSION OF METALLICITY UNCERTAINTIES FROM SEC8?

\subsection{Interstellar Reddening from the Literature} \label{ssec_samebv}

%\textbf{ [Material originally in this section was blended with text from the original appendix A.1, and the figure moved here]: }
Interstellar absorption, $A_K$, is a critical component of converting a star's apparent magnitude into its absolute magnitude.  
%For each star in our sample, we computed $A_K$ from the interstellar reddening using $A_K = 0.36 \times E(B-V)$ \citep{cardelli89}, having found $E(B-V)$ values for each star from various sources in the literature.  
In general, absorption is proportional to interstellar reddening, $E(B-V)$, and a variety of sources in the literature can yield reddening values for a given star.  These sources have improved markedly since \citet{fernley98}, who primarily used the reddening maps of \citet{bh78}, supplemented with estimates from $(V-K)$ colors for low-latitude stars.  To reduce the effects of interstellar absorption, we avoided inclusion of stars with low Galactic latitude unless they were of specific astrophysical interest (metallicity, period, etc). 

\citet{sfd98} used all-sky dust emission maps %produced using data 
from the {\it IRAS} and {\it COBE/DIRBE} satellites to estimate the interstellar reddening $E(B-V)$ as a function of Galactic coordinates.  We obtained an $E(B-V)$ value for every star in our sample and applied the recommended correction from the improved calibration of \citet{schlafly11}.  \citet{sfd98} estimated the uncertainty in their reddenings to be 16\% of the reddening value, which we adopted.  

We also compiled $E(B-V)$ estimates from \citet{blanco92}, who used a metallicity-dependent relation to predict the intrinsic $(B-V)_0$ color at minimum light.  Also, for every star in Table~\ref{tab_targets} with %existing infrared photometry in the column labeled $K_{F98}$, 
 $V$ and $K$ magnitudes from \citet{fernley98}, we calculated the observed $(V-K)$ color and estimated $E(B-V)$ following the procedure in Sec.~2.2 of \citet{fernley98}.  Specifically, we used Eqn.~9b of \citet{fernley93} along with the period and [Fe/H] from Table~\ref{tab_targets} to estimate each star's intrinsic $(V-K)_0$ color; we fit polynomials to the coefficients $c_3(Z)$ and $c_6(Z)$ in Tables~2 and 4a of \citet{fernley93} to smooth the behavior of the estimates.  \citet{fernley98} estimated the reddening uncertainty in this method to be 0.03 mag.  These three values of $E(B-V)$ are presented in Table~\ref{tab_targets} as $E_{SF11}$, $E_{B92}$ and $E_{F98}$, respectively.

\begin{figure}
%aastex52 \epsscale{.80}
%\plotone{fp18_ebv.eps}
\plotone{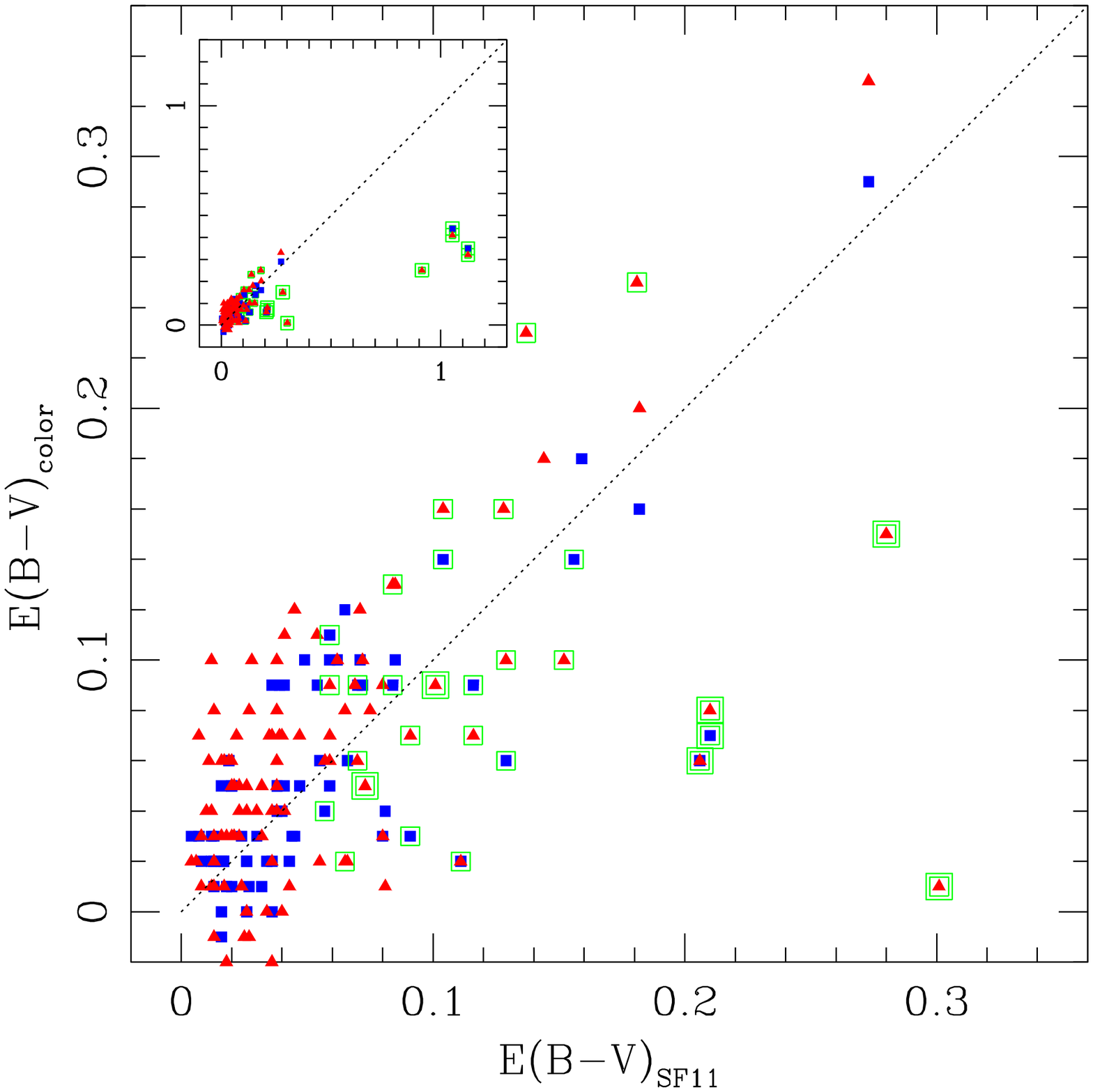}
\caption{Interstellar reddening values based on $(B-V)$ colors \citep{blanco92} [solid squares] and $(V-K)$ colors \citep{fernley98} [solid triangles] are plotted against those from the dust map recalibration of \citet{schlafly11}.  Stars with $|b| < 20^\circ$ are marked with a  single open square, and stars with $|b| < 10^\circ$ are marked with a  double  open square.  The dotted line marks the 1:1 relationship.  The inset  includes stars with large reddenings in our  working sample of 172 stars. \label{fig_ebv}}
\end{figure}

Figure~\ref{fig_ebv} inter-compares these values.  For stars with $E(B-V)_{SF11} \lesssim 0.2$ mag, both color methods follow the 1:1 correlation line with no obvious systematic offsets, so we can combine them without correction. %MOVE THIS PHRASE TO SEC7.  
The rms scatter of the Blanco values around the 1:1 line is 0.030 mag, while the rms is 0.034 mag for the $(V-K)$-based values.  Stars with larger reddening, which tend to be close to the Galactic plane, have color-based reddenings that are smaller than predicted by the dust-maps, suggesting the dust columns may extend beyond the distance of these stars.
%which are known to be unreliable at low Galactic latitudes \citep{schlafly}(??).   "Untested at |b| < 10" sfd98.
%NEED A CONCLUDING SENTENCE, OR COMBINE REDDENINGS TO AK HERE NOT SEC 7.  
In general, the use of three separate methods for finding the interstellar reddening and absorption enables us to detect and reject outliers, and establish a more representative mean value. In Sec.~\ref{ssec_meanak} we describe and apply a weighting scheme to combine these values into a single value for the interstellar absorption for each star; we apply a similar scheme to the stars' photometric values as well.

\subsection{Photometry from the Literature}

The $K_{F98}$ column in Table~\ref{tab_targets} lists the multi-epoch $K$-band mean magnitudes available in the   literature compiled by \citet{fernley98}.  Our photometry reference numbers 1--3 in this table match those of \citet{fernley98}: reference number 1 indicates the star is among the 29 listed in Table~16 of \citet{skillen93} for which dozens of observations, distributed uniformly in phase, were acquired as part of previous Baade-Wessilink (BW) studies   by three separate teams 
%(we use a hybrid reference number in Table~\ref{tab_targets} in which the leading "1" indicates the citation stream from \citet{fernley98} and \citet{skillen93}, and the trailing decimal portion 
(a note in the Comments column cites the source of the original photometry, as identified in the notes to Table~\ref{tab_targets}).   The density of points in the light curves of these stars makes them the most reliable stars in our study.  Reference number 2 indicates that the $K_{F98}$ value is based on the mean of a handful of photometric observations by \citet{fernley93b}, while \#3 refers to unpublished photometry reported in \citet{fernley98}. 
%\textbf{ These first 75 rows in Table~\ref{tab_targets} list the stars we observed as part of this project. }

The recent compilations in Table~5 of \citet{monson17} and Table~1 of \citet{hajdu18}  support our literature search that found no new multi-epoch $K$-band data published since \citet{fernley98}, 
%(Hajdu Tab1 has 19 classic field RRL, missing AU and ??? Vir) 
with the exception of NSV~660 which was observed extensively by 2MASS in one of its calibration fields \citep{szabo14}.  Because the light curve of this star was obtained with the 2MASS $K_s$ filter, we add NSV~660 to our collection of observations in Table~\ref{tab_targets}.     \citet{szabo14} included a period and estimated $[\mathrm{Fe/H}] = -1.31 \pm 0.03$ dex for NSV~660 but noted the possibility of systematic uncertainties, so we adopt an uncertainty of 0.10 dex.  This star has since been named LP~Cet, but we retain the older name to emphasize its distinct nature from the other stars in our study.  
%; the period, metallicity, and apparent magnitude data for this star are from \citep{szabo14} [DOUBLE CHECK!]. 

% Moved to Sec.7.1: Both  \citet{monson17} and \citet{hajdu18} (see their Table~1) emphasized that archival data were obtained using  a mixture of photometric systems which may be less uniform than our $K_s$ data.  In their Table~5, \citet{monson17} provided intensity-mean values recalibrated to the 2MASS photometric system for twenty of the 29 BW RRL.
%%for which multi-epoch $K$-band data had been obtained in previous Baade-Wessilink (BW) studies (these stars have dozens of data points distributed uniformly in phase, making them the most reliable stars in our study; \citet{skillen93} list all 29 stars observed as part of BW studies).  Their careful recalibration of those data to the 2MASS system should make them consistent with our data.  
%These data should be consistent with our $K_s$ photometry, an assertion which we test in Sec.~\ref{sec_compare}.  
  We add to our Table~\ref{tab_targets}, after the entries for the 75 stars which we observed and the entry for NSV~660, 
an additional 24 stars with extensive photometry from BW analyses,  SW~And through UU~Vir. 
%\textbf{ We add an additional 24 stars with extensive photometry from BW analyses, } SW~And through UU~Vir, to our Table~\ref{tab_targets} after the entry for NSV~660 and the 75 stars which we oberved.
% ; we observed the other three stars, and have substituted \textbf{Monson's} intensity-mean $K_s$ values for the original values \textbf{from \citet{skillen93} in} the $K_{F98}$ column of our table.  For each of these twenty stars, marked in our table with reference number 4, \citet{monson17} estimated the uncertainty in their mean $K_s$ magnitude to be 0.009 mag.  
The final 47 stars in our table  have  lower quality  $K$-band photometry from \citet{fernley98}.  We have not obtained new observations for these stars, but will explore their photometric recalibration and overall utility in Sec.~\ref{sec_compare}.        
%We will utilize these stars in Sec.~\ref{sec_gaia} to 
The recent intensity-mean $K_s$ values based on single-epoch photometry \citep{dambis09, dambis13} are described and compared with   our results and the multi-epoch photometry from Table~\ref{tab_targets} in Sec.~\ref{sec_compare}.

%We also add to our Table~\ref{tab_targets} a line for AU~Vir, whose original $K$-band data on the SAAO system we converted to the 2MASS system using Eqns.~6 of \citet{monson17}   MOVE TO COMPARE??

%We obtained $K_s$ photometry for the 75 stars listed above NSV~660 in Table~\ref{tab_targets}.  The 70 stars listed below NSV~660 in Table~\ref{tab_targets} have $K$-band photometry from \citet{fernley98}, though as indicated in Table~1 of  \citet{hajdu18}, they are compiled from a variety of sources utilizing a mixture of photometric systems which may be less uniform than our $K_s$ data.  Nevertheless, we will employ these data in Sec.~\ref{sec_gaia} to expand our RRL luminosity calibration.

% Note about Dambis09 moved to conclusions ... future work

Stars known to exhibit the Blazhko effect are noted in the comments column of Table~\ref{tab_targets}.  {\citet{jurcsik18} showed that a star exhibiting a well-defined Blazhko cycle in the $I$-band shows related brightness variations in the $K$-band of $\sim$0.1 mag, less than half the star's Blazhko cycle $I$-band amplitude (see their Figure 3).  While this effect is small, observations made at different points in a star's Blazhko cycle could lead to a biased estimate of the star's intensity-mean magnitude, so we have intentionally avoided observing Blazhko stars unless they are of other astrophysical interest.  

%\citet{jurcsik18} were perhaps the first to detect photometric variations in RRL known to exhibit the Blazhko effect.  Their data show the amplitude of the Blazhko cycle modulations to be $\sim$0.1 mag in $K_s$, less than half of that seen in the $I$-band.  While this e

\subsection{Characterizing the Sample} \label{ssec_samchar}

Figure~\ref{fig_zper} plots period versus metallicity for the stars in Table~\ref{tab_targets}.  The identification of stars as fundamental or first-overtone pulsators, RRab or RRc respectively, was taken from the literature and was based on the stars' periods, amplitudes, and light curve shapes.  The well-known trend between period and metallicity is apparent, as is the separation between RRab and RRc.  We selected our targets (the first 75 stars listed in Table~\ref{tab_targets}) to complement the period-metallicity distribution of the 30 existing high-quality light curves produced in the course of past BW analyses.
% (see Table~16 of \citet{skillen93} for a list of these stars).

\begin{figure}
%aastex52 \epsscale{.80}
%\plotone{fig_zper.eps}
%\plotone{fig1.eps}
\plotone{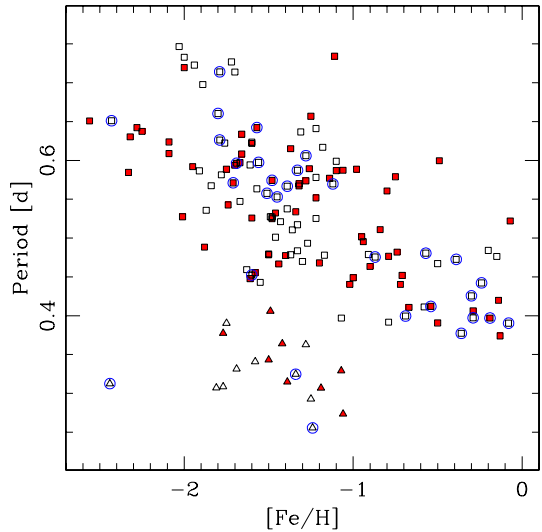}
\caption{Period is plotted against [Fe/H] for  the 146 RRL stars from Table~\ref{tab_targets}, with squares and triangles marking ab-type and c-type RRL, respectively.  Stars we observed are filled, and stars which have been observed as part of previous Baade-Wesselink studies are circled.
  \label{fig_zper}}
\end{figure}

%Using DR2 proper motions and radial velocity data from \citet{fernley98} and \citet{layden96}, we studied the $UVW$ space motions of stars in our database.
%(see the online Appendix for details).  

%\textbf{ [This paragraph was blended with text from the original appendix A.2 and the figure moved here]: }
We used the proper motions from the \textit{Gaia} DR2, and radial velocity data from \citet{fernley98} and \citet{layden96}, in the ``space motion'' solution method described in \citet{layden95} to obtain the three-dimensional velocities of the stars in our working database.  Figure~\ref{fig_uvw_feh} shows these $UVW$ motions plotted against metallicity.  The stars with higher metallicities have the smaller velocity dispersions and Sun-like Galactic rotation commonly found for the Galactic thick disk, whereas the more metal poor stars have the broad velocity dispersions and net rotation that lags the Sun by $\sim$200 km~s$^{-1}$ typical of stars in the Galactic halo \citep{layden95}.  The transition between these kinematic populations appears to be at [Fe/H] $\approx -0.9$ dex on our metallicity scale.
% \citep{gratton04}.  
%Note individual stars near *** Two of the metal-rich stars (*** at [Fe/H] = --0.9X and $U = YY$, and ** at [Fe/H] = --0.9X and $U = YY$) may be halo interlopers.  
Some of the stars with [Fe/H] $< -0.9$ that have low velocity dispersions and $V \approx 0$ km~s$^{-1}$ may be members of the metal-weak thick disk \citep{layden95,dambis09}.  We used an earlier version of this plot when selecting stars for our observation program to ensure a good distribution of stars around the disk-halo boundary, so that our data would be sensitive to any luminosity discontinuity between these two kinematic populations.
%Note individual stars from diskosity plot.  Put them in a table (Z, U, V, W, note)?

In total, Table~\ref{tab_targets} contains 147 stars that make up our initial data set, originally designed as a target list for \textit{SIM-PQ} but now useful in the era of \textit{Gaia}.

\begin{figure}
%aastex52 \epsscale{.80}
%\plotone{fig_uvw_feh.eps}
%\plotone{fig16.eps}
\plotone{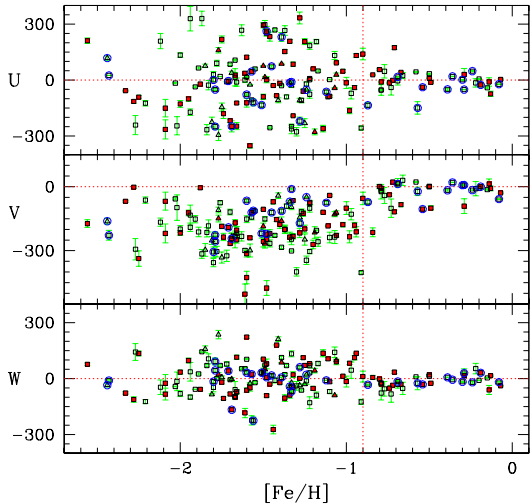}
\caption{The $UVW$ space velocities, in units of km~s$^{-1}$, are plotted against metallicity for RRL in our database, showing (top) the $U$ velocity component, directed towards and away from the Galactic center; (center) the $V$ velocity, along the direction of Galactic rotation; (bottom) the $W$ component, directed perpendicular to the Galactic plane.  Squares and triangles mark ab- and c-type RRL.  Filled symbols mark stars we observed in the $K_s$-band, while circles mark stars with high-quality BW photometry. \label{fig_uvw_feh} }
\end{figure}

% Pre-ref text from Sec.2, largely replace with Appendix text:
%We found that the stars with [Fe/H] $> -0.9$ dex on this metallicity scale have kinematics consistent with membership in the Galactic thick disk, while more metal poor stars show the large velocity dispersions of the Galactic halo, though it is likely that some of the stars in this group represent the metal-weak tail of the thick disk \citep{layden95,dambis09}.  
%Using the period-shift analysis described in the online Appendix, we separated the halo stars into short- and long-period groups which we tentatively associate with the Oosterhoff~I and II groups, respectively.  
%We mark these stars as ``OoI'' and ``OoII'' in the comments column of Table~\ref{tab_targets}.  
%In total, Table~\ref{tab_targets} contains 147 stars that make up our initial data set, originally designed as a target list for \textit{SIM-PQ} but now useful in the era of \textit{Gaia}.

\bigskip

\section{Observations}  \label{sec_obs}

%The standard approach to variable star photometry is to obtain contemporaneous photometry in two or more filters so that instantaneous colors can be obtained that enable determination of the magnitude in each bandpass.  Because we required only $K$-band photometry, and to maximize efficiency during our limited observing runs, we employed a different approach using only the $K_S$ filter.  We describe the method in Sec. \ref{sec_calib} and its effect on our photometry in Sec. \ref{sec_errs}.  Since our preferred targets include stars spread across the entire sky, we obtained images of these RRL  using two different sets of equipment. 

  Since our target stars shown in Table~\ref{tab_targets} are spread across the entire sky, we obtained images of these RRL  using two different sets of equipment. 
%\subsection{Southern Hemisphere}
%  120 nights in 2004A + 109 in 2004B = 229 nights
In the Southern hemisphere, we  used the 1.3-m telescope of the Small and Medium Aperture Research Telescope System (SMARTS) consortium located on Cerro Tololo in Chile in queue-scheduled mode during 229 nights between 2004 January 27 and 2005 January 30.  We used the ANDICAM instrument\footnote{Details are available at \url{http:www.ctio.noao.edu/noao/node/153}.} to obtain images simultaneously through the $K_s$ and Cousins $I_C$ filters using different optical paths and detectors within the instrument.
% using the infrared and optical paths, respectively.  
In this work, we used the $I_C$ images only for differential photometry that enables us to determine the phases of our $K_s$ images; a full analysis of the optical images will be reported in a forthcoming paper \citep{layden18}.  The infrared array was a Rockwell HgCdTe device with 18-micron pixels, operated with $2 \times 2$ pixel binning to give images with a $2.4 \times 2.4$ arcmin field of view and 0.27 arcsec pixel$^{-1}$ scale.  The north-east quadrant of the array was not functional so our images have an L-shaped field of view. %further restricting access to comparison stars.  
Exposure times of individual images were typically 6-20 s at each of five dither positions.  As part of the queue-observing process, calibration frames were obtained at regular intervals and these were applied to the raw images using standard procedures.  For a given star, the individual dithered images from a particular visit were combined to produce a single $K$-band image for that epoch.  The median seeing in this data set was 0.8 arcsec fwhm; the best and worst seeing values were 0.6 and 1.8 arcsec, respectively.

% The optical CCD is a Fairchild~447 $2048 \times 2048$ device with 15-micron pixles, read out with one (or two???) amplifiers.  Operated with $2 \times 2$ pixel binning, the images have a 0.37 arcsec pixel$^{-1}$ scale and a $\sim6 \times 6$ arcmin field of view.  Standard image processing including overscan correction, bias  subtraction, and division by a normalized sky flat field image was performed as part of the queue-observing procedure before the images were retrieved from the SMARTS archive.
% gain=2.3 e/adu, ron =6.5e, linear to <1% below 45Kadu (sat at 65K).

% ref or footnote?  \url{http:www.ctio.noao.edu/noao/node/153}

%\subsection{Northern Hemisphere}

  In the Northern hemisphere, we obtained images using the McGraw-Hill 1.3-m telescope at the Michigan-Dartmouth-MIT (MDM) Observatory on Kitt Peak in Arizona using the TIFKAM infrared imager during observing runs on 2006 June 9-25 and on 2007 June 23 through July 5.  We operated the telescope at $f/7.6$ so the $1024\times1024$ HgCdTe array yielded images with a $6.6\times6.6$ arcmin field of view and 0.38 arcsec pixels. Images of most target stars involved a sequence of ten 10- or 20-s exposures through the $K_s$-band filter, coadded on the array.  Typically, each visit to a star involved four such images, slightly dithered between exposures to shift the stars onto different sets of pixels.  These images were dark-subtracted and flat-fielded using dome flats, and the four images of each target star where shifted and combined using bad-pixel rejection into a single image per visit.  Typical seeing on the MDM images was 1.6 arcsec fwhm ($\sigma = 0.2$ arcsec).  

%Review notes on image processing and include (input from Glenn?).
% Mention corrections, V-corr in 06 and linear in 07??  Do this here or in sec_calib???

For both the Northern and Southern observing campaigns, we planned the observing of each star using its known period to obtain phase coverage that was as uniform as possible.  We aimed to get twenty observations per star for the southern, SMARTS targets, and owing to the more constrained observing time in the north, ten observations per star observed from MDM Observatory.  In practice, weather and timing constraints %often 
  sometimes affected our ability to attain these   goals: the median number of observations per star that we secured were 17 using SMARTS and ten using the MDM facilities.

\section{Photometry and Calibration}  \label{sec_calib}

%\textbf{ [Moved from start of Sec. 3:] } 
The standard approach to variable star photometry is to obtain contemporaneous photometry in two or more filters so that instantaneous colors can be obtained that enable determination of the magnitude in each bandpass.  Because we required only $K$-band photometry, and to maximize efficiency during our limited observing runs, we employed a different approach using only the $K_S$ filter   to perform differential photometry with respect to on-frame comparison stars of known magnitude and color.   We describe the details of the method below and discuss its effect on our photometry in Sec. \ref{sec_errs}.  

Because our target RR~Lyrae stars were selected to be of low-reddening, hence far from the Galactic plane, crowding was rarely an issue, and we performed aperture photometry using the implementation of DAOPHOT \citep{stetson87} in IRAF to measure instrumental photometry of the variable and comparison stars on each image.  On most images, we chose a stellar aperture with a diameter 
% form MDM, of 5.3 arcsec, about ... (variations in stellar profile from MDM visible when the seeing was good)
3.5 times the typical seeing. 
%\textbf{ [Removed some detail here about variations in stellar profile seen in best-seeing MDM images] }
%, large enough to capture most of the light from the star so that any small variations in the stellar profile  would be contained within the aperture.  

% For MDM data, used readnoise=5.5e, gain=4.0e/adu, datamax=10,000adu and a 6.5-7.0 pix aperture.

We selected comparison stars in the field of view of each variable star and retrieved their photometry from the 2MASS Point Source Catalog \citep{skrutskie06}.  We selected stars of similar brightness and color to the variable, finding  between one and twelve (typically four) suitable comparison stars in the various fields of view (the smaller, L-shaped field of view of the SMARTS images tended to permit fewer comparison stars, typically two).  We assumed that our instrumental magnitudes, $k$, would be related to the standard 2MASS magnitudes, $K_s$ by a function with the form
\begin{equation}
k - K_s = c_0 + c_1(J-K_s),  \label{eqn_calib}
\end{equation}
where $(J-K_s)$ is the standard color from the 2MASS list.  We performed least-squares regressions on data for comparison stars from a number of the richer fields to determine the values of the coefficients $c_0$ and $c_1$.  While the zero-point $c_0$ varied from image to image as the airmass and transparency varied, the color-term $c_1$ tended to remain stable, so we averaged the results to obtain $c_1 = +0.024 \pm 0.015$ from 21 fields for the MDM data.  This approach was complicated for the SMARTS data 
%because the smaller field of view allowed fewer 
by the smaller number of on-frame comparison stars; we obtained $c_1 = -0.002 \pm 0.051$ from 14 fields, a value consistent with the experience of the SMARTS technical team, who found found no discernible color-term using this equipment \citep{pogge18}.  Systematic effects resulting from these color-terms are addressed in Sec. \ref{sec_errs}.

% Note on smallness and matching of filter and array QE to std system?

% Email from Rick Pogge (2018mar30): Yale unable to find a significant color term (c1 ~ 0) and not able to fit a c0 value better than Frogel98 0.085.  Typical zero-pt uncertainties are ~0.05 mag in the IR.  Good results w/o color term.

On each image, we used the instrumental magnitude of the variable star, $k_v$ and one of the comparison stars, $k_c$, along with the 2MASS magnitude and color of the comparison star ($K_c$ and $(J-K)_c$, respectively) to estimate the standard magnitude of the variable star, $K_v$, differentially using the equation
\begin{equation}
  K_v = K_c + k_v - k_c - c_1[ (J-K)_v - (J-K)_c ].  \label{eqn_diffl}
\end{equation}
The 2MASS data includes values for $K_v$ and $(J-K)_v$, taken at a single epoch for each variable star.  We used the latter to calibrate our differential photometry for each variable star using Eqn. \ref{eqn_diffl}.   In reality, the surface temperature and hence the color of the variable changes as the star pulsates;  the $J-K$ color range is $\sim$0.25 mag for a typical ab-type RRL, and $\sim$0.1 mag for a typical c-type \citep{skillen93}.  
We discuss in Sec. \ref{sec_errs} how this simplification affects our photometric calibration, and we describe  how the 2MASS values for the RRL and a check star are used to test our calibrated light curve for each variable star.  

On each image, we used Eqn. \ref{eqn_diffl} to calculate $K_v$ for the variable from each available comparison star, leading to $N_{\rm comp}$ separate estimates of the RR~Lyrae's brightness.  We took the mean of these $N_{\rm comp}$ values, giving double-weight to stars that were bright or particularly close in color to the variable.  The standard error of the mean of these $N_{\rm comp}$ values, $\sigma_m$, provides an uncertainty estimate for the mean value, $K_s$,   for this observation.   These data are presented in Table \ref{tab_timeser} for each image, where the columns include the star name, heliocentric Julian Date of the observation, the pulsation phase (see Sec. \ref{sec_tmplt}), the airmass of the star at the time of observation, and the observatory from which the image was acquired.  The last column provides an integer code describing the quality of the variable star's image profile estimated by visual inspection of each image: $Q = 2$ indicates a normal, high quality profile, while $Q = 1$ flags a lower quality profile in which trailing due to telescope drift, interference from a flat-fielding artifact, or proximity to the edge of the fully-calibrated region of the image makes the $K_s$ value somewhat less certain than its $\sigma_m$ value may suggest.  Of the 1060 photometric measurements reported in Table~\ref{tab_timeser}, only 62 have low quality.

%We repeated this calculation for each of the $N_c$ comparison stars
%visible on that image to arrive at $N_c$ separate estimates of the
%variable star's magnitude.  We repeated this procedure for each image to obtain an array of magnitudes for KIC across up to 17 comparison stars on each image of the time series (102 in $V$ and 559 in $I_C$).

\begin{deluxetable*}{lrrrrrccc}
%\tabletypesize{\small}
%aastex52 \tabletypesize{\scriptsize}
%aastex52 \rotate
\tablecaption{Time-Series Photometry of the Variable Stars \label{tab_timeser}}
% tab
%aastex52 \tablewidth{0pt}
\tablehead{
\colhead{Star} & \colhead{HJD} & \colhead{$\phi$} & \colhead{$K_s$} &  
\colhead{$\sigma_m$} & \colhead{$N_{\rm comp}$} &  \colhead{Airmass}  & 
\colhead{Obs\tablenotemark{a}} & \colhead{$Q$\tablenotemark{b}}
%& \colhead{$Q$\tablenotemark{b}}  
%\colhead{Image~Name} & 
}
%\colnumbers
\startdata
DM~And  & 2453898.9050 & 0.717 & 10.616 & 0.006 & 5 & 1.48 & 2 & 2 \\
DM~And &  2453898.9504 & 0.789 & 10.672 & 0.010 & 5 & 1.21 & 2 & 2 \\
\nodata & \nodata & \nodata & \nodata & \nodata & \nodata & \nodata & \nodata  & \nodata  \\
DM~And  & 2454285.8761 & 0.578 & 10.531 & 0.006 & 6 & 1.29 & 2 & 2 \\
WY~Ant  & 2453035.7338 & 0.388 & 9.535 & 0.020 & 1 & 1.02 & 1 & 2 \\
WY~Ant &  2453038.6888 & 0.533 & 9.580 & 0.020 & 1 & 1.08 & 1 & 2 \\
\nodata & \nodata & \nodata & \nodata & \nodata & \nodata & \nodata & \nodata  & \nodata  \\
AV~Vir  & 2453906.7785 & 0.831 & 10.710 & 0.004 & 3 & 1.82 & 2 & 2 \\
 \enddata
\tablenotetext{a}{An integer code describing the observatory utilized: 1 = SMARTS, 2 = MDM.}
\tablenotetext{b}{An integer code describing the quality of the variable star's image profile and hence photometric reliability: 1 = low, 2 = high.}
\tablecomments{(This table is available in its entirety in machine-readable form.)}
%\tablecomments{Table \ref{tab_timeser} is published in its entirety in the electronic edition of the \textit{Astronomical Journal}. A portion is shown here for guidance regarding its form and content.}
\end{deluxetable*}
% table2_timeseries.txt and selected table entries above DONE 2018may09.

\section{Template Fitting}  \label{sec_tmplt}

Our primary goal is to obtain the intensity-mean $K_s$ magnitude of each star in Table \ref{tab_targets}.  To accomplish this goal, we folded the observed data from Table~\ref{tab_timeser} for a given star with the star's known period, and fit the resulting phased light curve with a sequence of light curve templates. 
%In most cases?  
Following \citet{layden98}, the fitting adjusted each template by shifting it vertically to obtain a magnitude zero-point, shifting it horizontally to match the phase of the observations, and stretching it vertically to obtain the amplitude, using the downhill simplex method \citep{numrec} to minimize the error-weighted $\chi^2$ between the observations and the template.  We did a three-parameter fit for each template in the sequence, inspected visually the four best fits, and in the vast majority of cases we adopted the fitted template with the smallest $\chi^2$.  

%\textbf{ [Moved from Sec. 2:] 
In addition, we fit our observations with both of the periods shown in Table~\ref{tab_targets} 
%(from VSX and \citet{fernley98}) 
and inspected the resulting phased light curves in both $K_s$ and $V$ or $I$.  In cases where the VSX period produced light curves with less scatter, or when they were indistinguishable, we adopted the VSX period: it is reported in Table~\ref{tab_finK} along with the reference code 1.  In cases where the period from \citet{fernley98} was preferable, we adopted it and report it in Table~\ref{tab_finK} with reference code 2.  For four stars, neither period adequately phased the observations, so we obtained an improved period using our optical data (reference code 3; see details in the Appendix). We also include in Table~\ref{tab_finK} an epoch associated with each star's maximum in optical brightness. } 
%Because the light curve shape is a strong function of wavelength, we require templates suited to the infrared.

% \bibitem{Press et al.(1986)} [numrec] Press, W. H., Flannery, B. P., Teukolsky, S. A., and Vetterling, W. T. 1986, Numerical Recipes: The Art of Scientific Computing, Cambridge Univ. Press, Cambridge, p.289

\subsection{Templates in $K$}

% \bibitem{Jones et al.(1996)}[jcf96] Jones, R. V., Carney, B. W., \& Fulbright, J. P. 1996, \pasp, 108, 877  (JCF96)

Because the light curve shape is a strong function of wavelength, we require templates suited to the infrared.  
\citet{jcf96} developed template RRL light curves from observed $K$-band observations of 17 ab-type and four c-type stars (see their Table 3) by fitting Fourier series to the observed light curves after grouping them by $B$-band amplitude, and hence light curve shape (see their Figure~1).  We used their Fourier coefficients (see their Table~4) to construct templates, named ``ab1'' thorough ``ab4'' and ``c,'' sampled at intervals of 0.02 phase units yielding 51 phase points.  We derived an additional template, named ``ab5,'' by fitting a smooth curve to the data in their Figure~1f in order to capture the light curve shape with the sharpest peak.

%We normalized the templates to an amplitude of 1.0 and shifted them vertically so minimum light fell at a magnitude of 0.0 in order to prepare them for the template-fitting code used in \citet{layden98}.  We also 
We shifted these templates in phase so that the reference phase, defined by the sharp minima in the RRab light curves, fell at the
%theiir natural 
  phases shown in Figure~1 of \citet{jcf96} that were based on the stars' optical ephemerides (also see their Figure~2) rather than at $\phi = 0$ as defined by the Fourier coefficients shown in Table~4 of \citet{jcf96}.   This has the advantage that when the template is fitted to observed $K_s$ data, $\phi = 0$ should correspond to the maximum brightness in the $V$-band.  

\citet{jcf96} discussed a use of their templates in which the ephemeris of an RRL is known but only 1-2 $K$-band observations have been obtained.  They developed phase-amplitude (their Fig.~2 and Eqn.~6) and amplitude-amplitude (their Fig.~3 and Eqns.~7-9) relations that would enable a star's $B$-amplitude value to predict the star's phase and $K$-amplitude, so the appropriate template could be matched to the $K$-band observation to obtain the star's mean $K$-band magnitude.  We chose to gather more than 1-2 observations per star for several reasons. %First, we did not have optical photometry WE DID!  USE BGSU DATA!
Our chief concern was the degree of scatter and bifurcation of their $B$- versus $K$-amplitude relation, which could result in the use of the wrong amplitude for a given star.  Second, each template in \citet{jcf96} is derived from the observed light curves of 3-7 individual stars %, grouped by $A(B)$ amplitude 
(see their Table~3); these stars may not represent the complete range of RRL light curve shapes as seen across a wider range of metallicities and stellar populations (for example, see the cases of RV~Cap and RX~Col below).  Finally, by obtaining more data points and fitting a template, we reduced the effect of random photometric errors in each observation.  However, as described below, we did perform such one-parameter fits for several stars whose observations were insufficient for our preferred three-parameter fitting approach.

\subsection{Fitting SMARTS data}

\begin{figure}
%aastex52 \epsscale{.80}
%\plotone{fig_lc1.eps}
%\plotone{fig2.eps}
\plotone{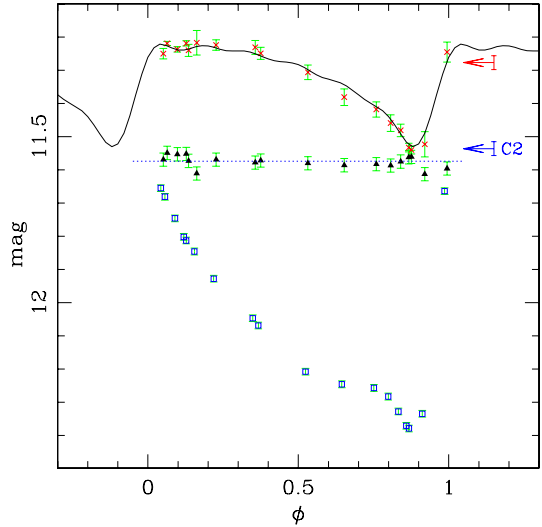}
\caption{The phased light curves of BP~Pav: crosses show our SMARTS $K_s$ magnitudes and the solid curve is the template fit to them.  The triangles are our $K_s$ magnitudes of the comparison star C2, and the dotted line shows their median magnitude.  The horizontal arrows show the 2MASS $K_s$ magnitude of BP~Pav (upper) and of C2 (lower), and the error bars show their 2MASS uncertainties.  The open squares show contemporaneous $I$-band magnitudes of BP~Pav, shifted vertically for convenience of display; these data peak at a phase of 0.0 confirming that the $K_s$ data were correctly phased by the three-parameter template fit (see Sec. \ref{sec_tmplt}).  \label{fig_lc1}}
\end{figure}

%Most of the stars observed with the SMARTS telescope in Chile 
Most of the southern stars observed with SMARTS were visited enough times so that one or more points were obtained during a star's steep brightness rise from minimum to maximum light.  This provided a good constraint on the horizontal, phase-shift component of the three-parameter fit for that star.  We tested whether the result was correctly phased by plotting the contemporaneous $I$-band photometry for each star using the same period and epoch as the $K$-band data, and in almost every case, the maximum in the $I$-band light curve occurred at a phase of 0.0 as expected for a correctly-phased $K$-band light curve.  Figure \ref{fig_lc1} shows an example of one fit.  Of the 42 stars observed with SMARTS, only two (RX~Cet and W~Tuc) failed to produce reliable three-parameter fits; their cases are described below and in the Appendix. 

We computed the intensity-mean $K_s$ magnitude for each star from the individual data points and from the best-fit template.  
%In most cases, 
  For stars with light curves that were well-sampled in phase, they differed by only a few millimag, so we averaged the values; in under-sampled cases we used the result from the template fit alone.   In most cases, we determined the light curve extrema, $K_{max}$ and $K_{min}$, from the best-fit template, since the scatter of individual data points can lead to spuriously large amplitudes.  The weighted standard deviation of a star's points around the best-fit template, $\sigma_{fit}$, is a useful measure of the quality of the fit.  This scatter is due to factors such as the photometric uncertainty in each observed data point (which is related to signal strength, sky noise, and errors in image processing), the horizontal scatter in points due to an error or change in a star's period (in most cases this was negligibly small because of the high precision of our periods and short interval of observation epochs), and any cycle-to-cycle modulation of the RRL light curve due to intrinsic sources like the Blazhko effect.  For the forty stars with three-parameter fits, the median value of $\sigma_{fit}$ was 0.027 mag.
% ave=0.032 stdev=0.016

%  Fig 5 BP Pav went here originally

\subsection{Fitting MDM data}  \label{ssec_mdmfit}

Of the 34 northern stars observed with MDM, 28 had enough data points with adequate phase distribution to constrain reliable three-parameter light curve fits.  In general, these stars had a larger number of on-field comparison stars than the SMARTS data, and the uncertainties in their individual $K_s$ magnitudes were smaller, resulting in a median $\sigma_{fit}$ of 0.013 mag.  
% ave=0.014 stdev=0.005
These stars did not have contemporaneous optical data with which to confirm their phases, but for each star we have uncalibrated $V$ or $I$ time-series photometry from the 0.5-m telescope at Bowling Green State University (BGSU) in Ohio, which will be presented in a forthcoming paper \citep{layden18}. These optical data were often acquired several years before or after the MDM data, so small errors or changes in a star's period could result in a phase offset between the optical and $K_s$ data.  We therefore fit the light curve with both the %original 
  \citet{fernley98} and VSX periods  %(see Sec.~\ref{sec_sel}; for four stars, we obtained an improved period from our optical data) 
and adopted the period giving the tightest light curves and the maximum at optical light closest to $\phi = 0$.  While the choice of periods made little difference in the distribution of $K_s$ points in phase (the MDM data were taken over a period of time -- days or up to one year -- that is short compared with the time between when the MDM and BGSU data were taken), in some cases it resulted in a small phase shift and consequent selection of a different best-fit template.  The difference in the resulting mean $K_s$ magnitude was always insignificant (mean $= -0.001 \pm 0.003$ mag).    As a check on the size of phase offsets that might occur during the time elapsed between the acquisition of the optical and infrared images, we computed the number of cycles elapsed using the adopted period, and one computed using a reasonable offset in that period (based on the published precision of the adopted period).  Only two stars had unacceptably-large cycle-count discrepancies (XZ~Dra and RV~UMa; see the Appendix for details), while the rest had discrepancies smaller than $\sim$0.02 cycles.  
For all of the 28 stars with three-parameter fits, visual inspection of the light curves led us to a high confidence in the quality of the fits.  The mean and extreme $K_s$ values for these 28 stars were calculated as described in the previous subsection for the SMARTS stars.

% \bibitem{park13} Park, J.-H., Lee, J. W., Kim, S.-L., Lee, C.-U., and Jeon, Y.-B. 2013, Publ. Astron. Soc. of Japan, 65, 
% Good references to history of slow realization that BX Dra was not RRL but WUMa --  https://doi.org/10.1093/pasj/65.1.1

\bigskip
\bigskip
\bigskip

\subsection{Other Fits}

For six of the MDM stars and two of the SMARTS stars, the $K$-band observations were insufficient in number or phase distribution to adequately constrain the fits.  For each of these stars, we made a preliminary phased light curve using an arbitrary epoch and the known period for both the $K$-band and optical data, and determined a phase shift that would bring the peak of the optical light curve to $\phi = 0$, thereby phasing correctly the $K_s$ data.  For cases where the   \citet{fernley98}  and VSX periods produced significantly different light curves, we again  adopted the period yielding the tightest light curve and the optical-light maximum closest to $\phi = 0$.  For one star, YZ~Cap, constraining the phase this way was sufficient to obtain an acceptable two-parameter fit which solved for the star's amplitude and zero-point (see the Appendix for details).

\begin{figure}
%aastex52 \epsscale{.80}
%\plotone{fig_lc2.eps}
%\plotone{fig3.eps}
\plotone{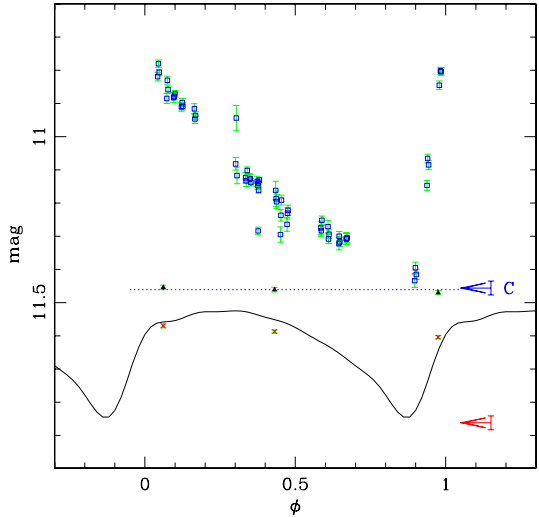}
\caption{The phased light curves of AA~Aqr.  The symbols are as in Figure~\ref{fig_lc1}.  Comparison star ``C'' is shown as a non-variable check star.  The $I$-band observations were obtained at BGSU less than one year before the MDM $K_s$ observations, and their maximum was shifted to $\phi = 0$ to establish the phases of the $I$ and $K_s$ data.  The template ``ab2'' was scaled to an amplitude of $\Delta K = 0.32$ mag based on the $V$-band amplitude of \citet{munari14}, and shifted vertically using a one-parameter fit.  \label{fig_lc2}}
\end{figure}

For the seven remaining stars, there were too few points in the critical phases around the deep, brief minima to adequately constrain the amplitude of the fit.  Here, we reverted to the original template fitting scheme outlined in \citet{jcf96}.  We obtained a star's optical amplitude from the literature, often $B$-band data from \citet{bookmeyer77}, and used Eqn.~7 of \citet{jcf96} to predict the star's $K$-band amplitude.  We also used the star's $B$-band amplitude to select a template from the groupings in Figure 1 of \citet{jcf96}, which demonstrates the correlation between amplitude and light curve shape.  We then performed a one-parameter fit using this template, amplitude, and phase-shift in order to find the zero-point that gave the best fit between the observations and scaled template.  Figure~\ref{fig_lc2} shows the fit for AA~Aqr, in which $I$-band photometry from BGSU was used to constrain the phase of the MDM $K_s$ photometry.  For the stars utilizing one- and two-parameter fits, mean and extreme $K_s$ magnitudes were obtained from the best-fit template alone,   and details of the fitting procedures are provided in the Appendix. 
%, since the observed points did not adequately 

%RX~Cet:  The intensity means of the three-, two- and one-parameter fits were $\langle K_s \rangle = 10.147$, 10.152, and 10.159 mag, respectively.   We adopt the results of the one-parameter fit.

% Fig 6 AA Aqr went here originally

%Some of the MDM stars, particularly with those with fewer than ten observations, the phase distribution did not adequately constrain the phase-shift, the amplitude of the light curve, or either.  

%Most of the southern stars observed with SMARTS were observed enough times so that one or more points were obtained during a star's steep brightness rise from minimum to maximum light.  This provided a good constraint on the horizontal, phase-shift component of the fit for that star.  We tested whether the result was correctly phased by plotting the contemporaneous $I$-band photometry for each star using the same period and epoch as the $K$-band data, and in almost every case, the maximum in the $I$-band light curve occurred at a phase of 0.0 as expected for a correctly-phased $K$-band light curve.  Because the SMARTS data was

%We performed three-parameter fits on each star using the original period and the VSX period; usually they produced nearly identical results but in a few cases ?? 

%For most stars, we had two very similar estimates of the period, one from our original compilation of data (from GCVS/Fernley/Layden?) and a second from a recent inspection of the International Variable Star Index (VSX)\footnote{Accessed circa 2018 January from \url{https://www.aavso.org/vsx/index.php}.}.

\subsection{Resulting Photometry}

\begin{figure}
%aastex52 \epsscale{.80}
%\plotone{fig_lc3.eps}
%\plotone{fig4.eps}
\plotone{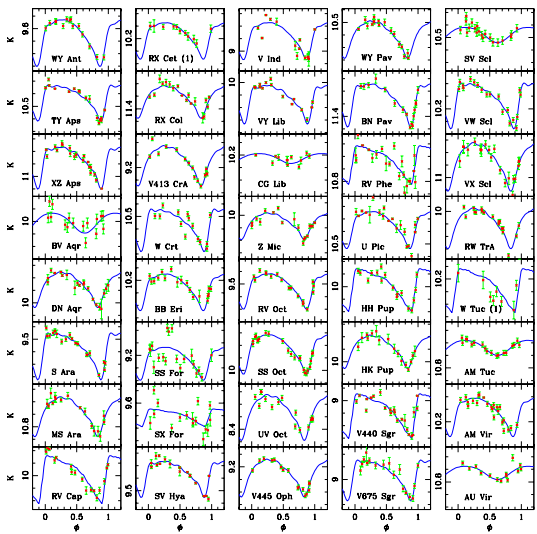}
\caption{The phased $K_s$ light curves of forty RRL observed with SMARTS.  All the templates were fitted using a three-parameter fit unless noted by ``(1)'' which indicates a one-parameter fit.  Each panel has a vertical range of 0.5 mag.  \label{fig_lc3}}
\end{figure}

\begin{figure}
%aastex52 \epsscale{.80}
%\plotone{fig_lc4.eps}
%\plotone{fig5.eps}
\plotone{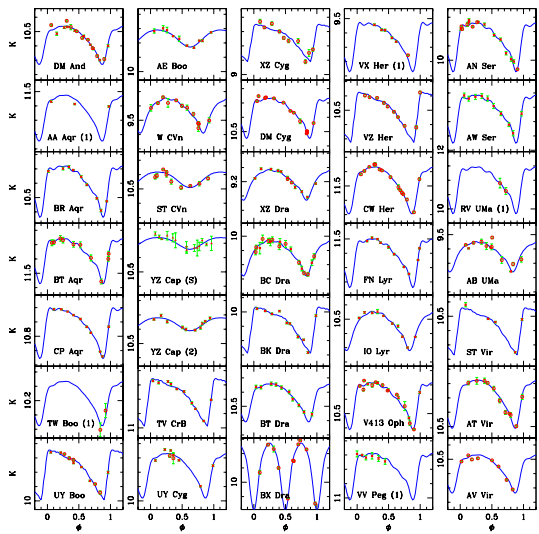}
\caption{The phased $K_s$ light curves of 34 stars observed at MDM.  Circles indicate 2006 data and crosses mark 2007 data.  All the templates were fitted using a three-parameter fit unless noted by ``(1)'' or ``(2)''  indicating a one- or two-parameter fit.  Each panel has a vertical range of 0.5 mag.  The star YZ~Cap was also observed with SMARTS; this light curve is indicated by the ``(S).''  The star BX~Dra was confirmed to be an EW-type contact binary system. \label{fig_lc4}}
\end{figure}

Figure~\ref{fig_lc3} shows our observed $K_s$ data for forty RRL observed with SMARTS and their template fits, while Figure~\ref{fig_lc4} shows the data and fits for 34 stars observed at MDM.\footnote{An individual light curve plot like those shown in Figures~\ref{fig_lc1} and \ref{fig_lc2} is available for each star at \url{http://physics.bgsu.edu/~layden/publ.htm}, along with other intermediate data products from this study.}   
Table~\ref{tab_finK} lists the intensity-mean magnitude, $\langle K_s \rangle$, for each star, along with its extrema $K_{max}$ and $K_{min}$, the number of observations $N_{obs}$, $\sigma_{fit}$ and the name of the best-fitting template, and the number of comparison stars $N_{comp}$ used to calibrate the differential photometry via Eqn.~\ref{eqn_diffl}.

  One star, BX~Dra, was listed by \citet{fernley98} as an RRab star and was thus included in Table~\ref{tab_targets}, but optical observations have shown  it to be a contact eclipsing binary system (see \citet{park13} and references therein).  Our $K_s$-band light curve shown in Figure~\ref{fig_lc4}  is consistent with this conclusion  and we do not consider BX~Dra further in the following analysis.

% Table~\label{tab_finK}
\begin{deluxetable*}{lccrlccrrrcccccc}
%aastex52 \tabletypesize{\scriptsize}
%aastex52 \rotate
\tablecaption{Final Photometry of the Variable Stars \label{tab_finK}}
% table4_sub2.txt --> tab4b.txt
%aastex52 \tablewidth{0pt}
\tablehead{
\colhead{Star} & \colhead{Obs\tablenotemark{a}} & \colhead{$N_{comp}$} & \colhead{$N_{obs}$} &  
\colhead{Period} & \colhead{Ref\tablenotemark{b}} & \colhead{Epoch} &
\colhead{$\langle K_s \rangle$} & \colhead{$K_{max}$} &  \colhead{$K_{min}$}  & 
\colhead{$N_{par}$} & \colhead{Temp\tablenotemark{c}} &
\colhead{$\sigma_{fit}$} & \colhead{$\sigma_z$} & \colhead{$\sigma_c$} & \colhead{$\sigma_K$}  
%& \colhead{$Q$\tablenotemark{b}
%\colhead{Image~Name} & 
}
%\colnumbers
\startdata
DM~And  & 2 & 6 & 18 & 0.630389  & 2 & 3898.45301 & 10.543 & 10.46 & 10.73 & 3 & ab2 & 0.016 & 0.007 & 0.001 & 0.008 \\
WY~Ant &  1 & 1 & 14 & 0.574312  & 2 & 3035.51097 & 9.639 & 9.53 & 9.90 & 3 & ab2 & 0.022 & 0.032 & 0.005 & 0.033 \\
\nodata & \nodata & \nodata & \nodata & \nodata & \nodata & \nodata & \nodata  & \nodata & \nodata  & \nodata  & \nodata  & \nodata   \\
AU~Vir  & 1 & 1 & 14 & 0.3432307 & 1 & 3077.67294 & 10.717 & 10.67 & 10.78 & 3 & c  & 0.026 & 0.030 & 0.019 & 0.036 \\
AV~Vir  & 2 & 3 & 8 & 0.6569073 & 1 & 3898.34958 & 10.548 & 10.47 & 10.74 & 3 & ab2 & 0.013 & 0.012 & 0.004 & 0.014 \\
NSV~660 & 3 & \nodata & 2969 & 0.636985 & 1 & 0659.80021 & 13.987 & 13.90 & 14.20 & 3 & ab2 & 0.074 & \nodata & \nodata & 0.02 \\
 \enddata
\tablenotetext{a}{The integer code identifying the observatory utilized: 1---SMARTS, 2---MDM, 3---2MASS.}
\tablenotetext{b}{Reference for the adopted period: 1---VSX, 2---\citet{fernley98}, 3---Determined in this work.}
\tablenotetext{c}{Name of the best-fitting template: ``ab1''--``ab5'' and ``c'' are from \citep{jcf96}, ``EW'' indicates a W~UMa-type contact binary star.}
\tablecomments{(This table is available in its entirety in machine-readable form.)}
%\tablecomments{Table \ref{tab_finK} is published in its entirety in the electronic edition of the \textit{Astronomical Journal}. A portion is shown here for guidance regarding its form and content.}
\end{deluxetable*}
% table2_timeseries.txt and selected table entries above DONE 2018may09.
%Star    TelNcoNobs <K>   Kmax  KminParTem sigf  sigz  sigc  sigK 

%\bigskip
%\bigskip

\section{Photometric Uncertainties}  \label{sec_errs}

It is important to obtain a reliable estimate of the uncertainty in the measured intensity-mean $\langle K_s \rangle$ magnitude for each star in our sample.  There are several sources of uncertainty.  First is the uncertainty in the vertical, zero-point level of our fitted template.  We noted earlier that $\sigma_{fit}$ describes the point-to-point scatter of the observed points around the best fit, and we adopt a value based on the standard deviation of the mean as the uncertainty in the zero-point of the fitted template, $\sigma_t = \sigma_{fit}~N_{obs}^{-0.5}$.

The second source of uncertainty is that of the 2MASS magnitude of each comparison star, $\sigma_i$\footnote{These errors are listed in the 2MASS Point Source Catalog as ``k\_msigcom,'' see \url{https://irsa.ipac.caltech.edu/workspace/TMP_Qm3I7V_16109/Gator/irsa/17533/dd_17533.html}. \citet{skrutskie06} state that errors between different scans across the sky are always less than 0.02 mag, and the online documentation (see \url{https://www.ipac.caltech.edu/2mass/releases/allsky/doc/sec2_2a.html}) suggests the zero-point error is 0.007 mag.  This small systematic error is included in k\_msigcom.  We also note that, with one or two exceptions, all the comparison stars used in this study had the highest photometric quality flags from this catalog.}.  Typically, $\sigma_i \approx 0.02$ mag, increasing for fainter stars. These errors are mainly random, due to errors in point-spread-function fitting, read noise, flat-field noise, etc. \citep{skrutskie06}, so for an RRL with multiple comparison stars, we combined the errors via $\sigma_z = ( \Sigma~ \sigma_i^{-2} )^{-0.5}$ to obtain a single estimate of the uncertainty in the external calibration.  The value of $\sigma_z$ for each star is listed in Table~\ref{tab_finK}.

%Systematically, he says errors between different scans (tiles) across the sky are always less than 0.02 mag, and the online documentation to the 2MASS Point Source Catalog\footnote{ See \url{https://www.ipac.caltech.edu/2mass/releases/allsky/doc/sec2_2a.html}. } suggests the zero-point error is 0.007 mag.

The third uncertainty source is the uncertainty in our color term, $c_1$, used to account for the color difference between variable and comparison stars in Eqn.~\ref{eqn_diffl}.  Across our sample, the median 2MASS color of the RRL was $(J-K_s) = 0.30$ mag with a standard deviation of 0.07 mag, while the median comparison star color was 0.53 mag with a broader standard deviation of 0.20 mag.  For each variable-comparison star pair, we computed the color difference $\Delta(J-K_s)$ between the two stars' 2MASS colors, and calculated the resulting systematic uncertainty in the star's mean $K_s$ magnitude as $\sigma_c = \sigma_{c1} \times \Delta(J-K_s)$, where $\sigma_{c1} = 0.051$ mag for the SMARTS data and 0.015 mag for the MDM data.  For each variable with multiple comparison stars, we computed the weighted mean of the individual $\sigma_c$ values, reported for each RRL in Table~\ref{tab_finK}.  The median values of $\sigma_c$ were 0.015 and 0.004 for the SMARTS and MDM fields, respectively.

An additional source of %systematic 
error in our $K$-band light curves originates with our non-standard procedure for calibrating the differential photometry, i.e., using only one filter rather than two filters, which would enable the calculation of a variable star's instantaneous color.  In Eqn. \ref{eqn_diffl}, we used for $(J-K_s)_v$ the 2MASS single-epoch color of the RRL rather than its instantaneous color.  We therefore made an %unknown systematic 
error in all the magnitudes for a given star.  This error can be estimated as $\sigma_n = c_1 \times \sigma_{jk}$, where $\sigma_{jk}$ is the standard deviation of the $(J-K_s)$ values a star experiences over the course of its pulsation cycle.  We determined $\sigma_{jk}$ for each of the four ab-type RRL (obtaining an average value of 0.073 mag) and for the one c-type RRL (0.045 mag) in \citet{skillen93}.  Thus, we expect that each ab-type RRL observed with MDM ($c_1 = +0.024$) should have an additional component to its uncertainty of $\sigma_n = 0.002$ mag, and each c-type RRL an added uncertainty of 0.001 mag.  The small color term for SMARTS observations ($c_1 = -0.002$ mag) makes this component of the uncertainty negligibly small.

We calculated each of these four sources of uncertainty ($\sigma_t$, $\sigma_z$, $\sigma_c$, and $\sigma_n$) for each star and %combined them through a quadratic sum 
added them in quadrature to obtain an overall systematic uncertainty in the intensity-mean $\langle K_s \rangle$ magnitude for that star, $\sigma_K$, whose values are listed in Table~\ref{tab_finK}.  The relative sizes of these uncertainties gives a sense of the relative strengths and weaknesses of each component of the procedure for each star, and across all the stars in our sample.  For stars observed with SMARTS, the median value of $\sigma_K$ was 0.029 mag, while it was 0.013 mag for the MDM stars.  Based on the scatter of observed points around the best fit templates, we estimate the uncertainty in the extrema values $K_{max}$ and $K_{min}$ in Table~\ref{tab_finK} to be about 0.02 mag for a typical star, though stars with larger values of $\sigma_{fit}$ are expected to have larger uncertainties in their extrema.

We performed two tests on our photometry of each star to identify potential problems with our procedure.  First, for each RRL we selected one comparison star and processed it with our variable star software,   using the remaining $N_{comp} - 1$ comparison stars to determine its time-series differential photometry.  In all cases, like those shown in Figures~\ref{fig_lc1} and \ref{fig_lc2} for comparison stars ``C2'' and ``C'' respectively, we obtained a flat run of magnitude with phase and a scatter comparable in size to the uncertainty estimates from our individual photometric measurements.  For example, the rms scatter of the points for the comparison star C2 in Figure~\ref{fig_lc1} was 0.018 mag, while the mean uncertainty was 0.020 mag  for this SMARTS star; for the check star in Figure~\ref{fig_lc2} the values were 0.008 and 0.005 mag, respectively  (MDM); and for the better-sampled MDM star FN~Lyr, the values were 0.007 and 0.005 mag.  Furthermore, the median of our $K_s$ magnitude estimates for each check star were %always 
  statistically consistent with the star's 2MASS $K_s$ mag measurement, as shown by the labeled arrows in Figures~\ref{fig_lc1} and \ref{fig_lc2}.

The second consistency test was to plot the single-epoch $K_s$ magnitude of the RRL from 2MASS on each star's light curve, and visually check whether it fell between the $K_{max}$ and $K_{min}$ range of the fitted template.  For almost all our RRL, this was the case (for example, see Figure~\ref{fig_lc1}).   For three stars, the 2MASS magnitude was formally outside the template range, but by an amount consistent with the uncertainty in the 2MASS magnitude (see Figure~\ref{fig_lc2}).  Only for the stars DM~Cyg and SX~For was the 2MASS magnitude significantly outside the template range (by 0.07 and 0.16 mag, respectively).  We recommend independent photometry of these stars.

% Sec.7
\section{Photometric Comparisons}  \label{sec_compare}

%In assessing the reliability of a set of photometry, it is useful to compare with other data sets.  
We have one star in common between the southern SMARTS and northern MDM data sets, YZ~Cap.  Unfortunately, this provides a poor comparison because the SMARTS data for this c-type RRL had a fairly large scatter around the best template (see Figure~\ref{fig_lc4}), $\sigma_{fit} = 0.026$ mag, and had only two comparison stars that were somewhat faint, leading to a large uncertainty in the intensity-mean magnitude, $\langle K_s \rangle = 10.293 \pm 0.033$ mag.  The MDM data had more comparison stars and a %consequently 
smaller overall uncertainty, yielding $\langle K_s \rangle = 10.359 \pm 0.012$ mag.  The difference of $0.066 \pm 0.035$ ($1.9\sigma$) hints that the SMARTS data may be systematically brighter than the MDM data, though the evidence is not compelling and there is little fundamental reason to think our differential photometry should suffer a systematic offset.
%is weak, certainly not compelling, evidence that the SMARTS data could be systematically brighter than the MDM data.

\subsection{Compiled Photometry} \label{ssec_compile}

  In order to compare our photometry with existing work from the literature, we assembled the available data in Table~\ref{tab_comb4}.  
The second column contains the intensity-mean magnitudes derived from the BW studies noted in Table~\ref{tab_targets}.
%[Moved from original p7:] } 
Both  \citet{monson17} and \citet{hajdu18} (see their Table~1) emphasized that most archival data were obtained using  a mixture of photometric systems which may be less uniform than our $K_s$ data.  In their Table~5, \citet{monson17} provided intensity-mean values recalibrated to the 2MASS photometric system for twenty of the BW RRL in our study,   including the star RR~Lyrae itself. 
These data should be consistent with our $K_s$ photometry, and appear in the second column of Table~\ref{tab_comb4} labeled $K_{BW}^\prime$, where the prime indicates the original magnitudes have been transformed to the 2MASS system.  %an assertion which we test below. 

     For ten additional BW stars, we used the relations of \citet{carpenter03}, several of which appear as Eqns.~4-6 of \citet{monson17}, to transform the published intensity-mean magnitudes onto the $K_s$ system, as shown in Table~\ref{tab_BW}.  In this table, the second column contains the intensity-mean $K$ magnitude listed in the paper cited in column~three, and the fourth column identifies the equation used in the transformation.  The transformed values for these ten stars also appear in the second column of Table~\ref{tab_comb4} (note that VY~Ser was observed by two separate teams; we averaged the transformed results to achieve the value shown in Table~\ref{tab_comb4}).  For all 30 BW stars we adopt photometric uncertainties of 0.009 mag following \citet{monson17}.   For convenience, columns 3--4 of Table~\ref{tab_comb4} repeat our $\langle K_s \rangle$ and $\sigma_K$ values from Table~\ref{tab_finK}.  

%: RS~Boo (9.445 mag, from bw24), UU~Cet (? mag, from bw35), SW~Dra (9.426 mag, from bw22), TW~Her (

% NEW TABLE 5
\begin{deluxetable*}{lrrrrrrrcrrcrrc}
%aastex52 \tabletypesize{\scriptsize}
%aastex52 \rotate
\tablecaption{Combined Photometry and Interstellar Absorption \label{tab_comb4}}
% table5_sub2.txt --> tab5b.txt
%aastex52 \tablewidth{0pt}
\tablehead{
\colhead{Star} & \colhead{$K_{BW}^\prime$} & \colhead{$\langle K_s \rangle$} & \colhead{$\sigma_{K_s}$} &  
\colhead{$K_{F98}^\prime$} &  \colhead{$\sigma_{F98}$} & \colhead{$K_{D13}$}  &  \colhead{$\sigma_{D13}$} & 
\colhead{Code$_K$\tablenotemark{a}}  &   \colhead{$\bar{\langle K \rangle}$} & \colhead{$\sigma_{\bar{K}}$} & 
\colhead{Flag\tablenotemark{b}} &  \colhead{$A_K$} & \colhead{$\sigma_A$} &  \colhead{Code$_A$\tablenotemark{c}} 
}
%\colnumbers
\startdata
SW~And   & 8.511 & \nodata & \nodata & \nodata & \nodata &  8.509 & 0.034 & 1001 & 8.511 & 0.009 & 0 & 0.017 & 0.003 & 111 \\
XX~And  & \nodata & \nodata & \nodata &   9.491 & 0.044  & 9.411 & 0.035 & 0021 & 9.442 & 0.027 & 0 & 0.013 & 0.003 & 111 \\
AT~And  & \nodata & \nodata & \nodata &   9.041 & 0.044 & 9.090 & 0.036 & 0021 & 9.070 & 0.028 & 0 & 0.044 & 0.005 & 111 \\
DM~And  & \nodata &  10.543 & 0.008 & \nodata & \nodata &  10.575 & 0.038 & 0201 & 10.544 & 0.008 & 0 & 0.029 & 0.005 & 100 \\
\nodata & \nodata & \nodata & \nodata & \nodata & \nodata & \nodata & \nodata  & \nodata  & \nodata & \nodata & \nodata & \nodata & \nodata & \nodata \\
AV~Vir   & \nodata &  10.548 & 0.014 & 10.611 & 0.044 & 10.566 & 0.037 & 0221 & 10.555 & 0.013 & 0 & 0.008 & 0.003 & 101 \\
BN~Vul  & \nodata & \nodata & \nodata   & 8.791 & 0.044 &  8.665 & 0.033 & 0021 & 8.710 & 0.026 & 0 & 0.153 & 0.008 & 911 \\
NSV~660  & \nodata  &  13.987 & 0.020 & \nodata & \nodata & \nodata & \nodata  &  0400 & 13.987 & 0.020 & 0 & 0.011 & 0.004 & 100 \\
\enddata
%\tablenotetext{a}{For stars with Baade-Wesselink analyses we adopt photometric uncertainties of 0.009 mag as per \citet{monson17}.}
\tablenotetext{a}{This four-integer code describing the sources of photometry is defined in Sec.~\ref{ssec_wmean}.}
%\tablenotetext{a}{The code identifying the sources of the photometry in the weighted mean $K$ magnitude (see Sec.~\ref{sec_compare}).  }
\tablenotetext{b}{A flag indicating whether the individual photometric measures scatter more widely than indicated by the weighted mean, $\sigma_{\bar{K}}$ where 1 or 0 means yes or no, respectively.}
\tablenotetext{c}{The code identifying the sources of the reddening values from Table~\ref{tab_targets} used to compute the weighted mean interstellar absorption $A_K$, is defined in Sec. \ref{ssec_meanak}.}
\tablecomments{(This table is available in its entirety in machine-readable form.)}
%\tablecomments{Table \ref{tab_timeser} is published in its entirety in the electronic edition of the \textit{Astronomical Journal}. A portion is shown here for guidance regarding its form and content.}
\end{deluxetable*}
\begin{deluxetable}{lrcl}
%aastex52 \tabletypesize{\scriptsize}
%aastex52 \rotate
\tablecaption{Additional BW Photometry  \label{tab_BW}}
%aastex52 \tablewidth{0pt}
\tablehead{
\colhead{Star} & \colhead{$K_{BW}$} & \colhead{Ref\tablenotemark{a}} & \colhead{Equation} 
}
%\colnumbers
\startdata
RS~Boo & 9.445 & bw24 & \citet{monson17} Eqn.~4 (CIT) \\
UU~Cet & 10.850 &  bw35 & \citet{carpenter03} Sec.~c (ESO)\tablenotemark{b}   \\
SW~Dra & 9.326 & bw22 &  \citet{monson17} Eqn.~4 (CIT)   \\
TW~Her & 10.217 & bw24  &  \citet{monson17} Eqn.~4 (CIT)   \\
SS~Leo & 9.933 & bw33  &  \citet{monson17} Eqn.~5 (UKIRT)  \\
V445~Oph & 9.241 &  bw33  &  \citet{monson17} Eqn.~5 (UKIRT)  \\
AR~Per & 8.532 & bw10  &  \citet{monson17} Eqn.~4 (CIT)   \\
RV~Phe & 10.721 & bw35  & \citet{carpenter03} Sec.~c (ESO)\tablenotemark{b}   \\
VY~Ser & 8.780 & bw24 &  \citet{monson17} Eqn.~4 (CIT)  \\
VY~Ser & 8.793 & bw33  &  \citet{monson17} Eqn.~5 (UKIRT)   \\
W~Tuc & 10.354 & bw35  & \citet{carpenter03} Sec.~c (ESO)\tablenotemark{b}   \\
\enddata
\tablenotetext{a}{These citations are defined in the Notes to Table~\ref{tab_targets}.}
%\tablenotetext{b}{The equation is $(K_s)_{2MASS} = K_{ESO} + (-0.044 \pm0.004) + (0.000 \pm 0.011)(J-K)_{ESO}$.}
\tablenotetext{b}{The equation is $(K_s)_{2MASS} = K_{ESO} + (-0.044 \pm 0.004)$ mag.}
%\tablecomments{(This table is available in its entirety in machine-readable form.)}
%\tablecomments{Table \ref{tab_timeser} is published in its entirety in the electronic edition of the \textit{Astronomical Journal}. A portion is shown here for guidance regarding its form and content.}
\end{deluxetable}
%  DONE 2019mar22

%\textbf{ [This paragraph moved from the original discussion of photometric differences and adapted:] } 
The stars in Table~\ref{tab_targets} with reference~\#2 have $K_{F98}$ values from \citet{fernley98} who calculated the simple mean of photometry taken at 2-4 epochs timed to provide a range of phases \citep{fernley93b}.  These magnitudes are on the CTIO/CIT system, so we corrected them to the 2MASS system using Eqn.~4 of \citet{monson17}, which brought them 0.019 mag closer to our $\langle K_s \rangle$ magnitudes.    %We estimated the uncertainty in each mean magnitude as $\sigma~N_{obs}^{-0.5}$, where $\sigma = 0.07$ mag is the estimated uncertainty in a single photometric observation \citep{fernley93b}.  
The remaining stars in Table~\ref{tab_targets} (with reference~\#3) were unpublished, so we have no knowledge of their nature, and so we applied no photometric correction; \citet{fernley98} reported their uncertainties to be 0.04 mag.  We list these data in the column of Table~\ref{tab_comb4} labeled $K_{F98}^\prime$, where the prime again signifies that some values have been transformed.

%\textbf{ [This paragraph moved from middle of differences discussion:] } 
\citet{dambis09} provided a valuable, parallel set of infrared data on RRL. Following a method outlined by \citet{feast08}, \citet{dambis09} phased each star's single-epoch photometry from 2MASS using an external ephemeris and fit a template at fixed phase and amplitude to obtain the intensity-mean magnitude. %,  $\langle K_{DF} \rangle$.  
The estimated uncertainty in a typical fit was 0.03 mag, and most stars had uncertainties in their 2MASS photometry of $\sim$0.02 mag, giving a combined uncertainty of 0.03-0.04 mag in the 
%$\langle K_{DF} \rangle$ 
mean magnitude of each star.   However, $\sim$8\% of   these  stars did not have reliable ephemerides, so   Dambis  did not phase and fit templates to these 2MASS data, and simply adopted the 2MASS magnitude at unknown phase.  The $K$-band amplitude of a typical ab-type RRL is 0.3 mag, so one expects a small but significant subset of the photometry of \citet{dambis09} to scatter broadly around the stars' true mean magnitudes.   A dozen stars were added to this set of magnitudes by \citet{dambis13},   and an individual uncertainty value for each star was provided.   The resulting apparent magnitudes were used for the analysis of the \textit{Gaia} DR1 RRL parallaxes by \citet{clementini17}, and for the DR2 parallaxes by \citet{muraveva18}.  The magnitudes and uncertainties are presented in Table~\ref{tab_comb4} as $K_{D13}$ and $\sigma_{D13}$ respectively.

\subsection{Photometric Differences} \label{ssec_diffs}

\begin{figure}
%aastex52 \epsscale{.80}
%\plotone{fig_comp3b.eps}
%\plotone{fig6.eps}
\plotone{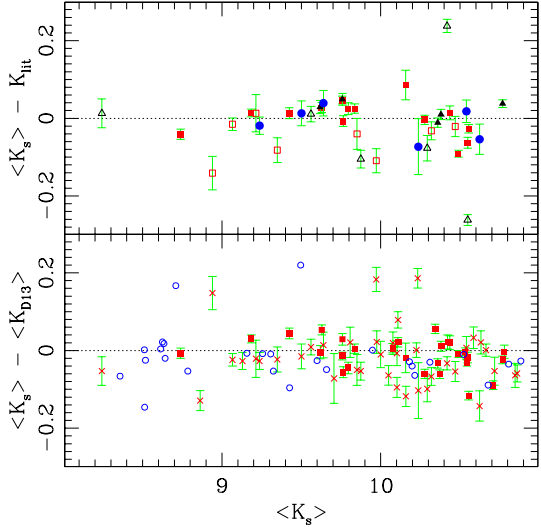}
\caption{Our intensity-mean $\langle K_s \rangle$ magnitudes are compared with photometry from the literature. The error bars show our photometric uncertainties, $\sigma_K$, alone. (Top) Circles mark six stars with densly-populated light curves (BW stars with photometry reference \#1 from Table~\ref{tab_targets}) after correction to the 2MASS photometric system; squares mark 22 stars from \citet{fernley93b} (reference \#2) after correction to the 2MASS system; and triangles mark ten stars with unpublished photometry reported in \citet{fernley98} (reference \#3).  Filled and open symbols indicate stars observed with MDM and SMARTS, respectively.    (Bottom) Our  $\langle K_s \rangle$ magnitudes from MDM (squares) and SMARTS (crosses) are compared with mean magnitudes from \citet{dambis13}.  The circles indicate photometry taken from the literature for the %29 
BW stars (reference \#1, after correction to the 2MASS system) compared with that of \citet{dambis13}.   \label{fig_comp}}
\end{figure}

Of the   74 RRL  we observed, 32 have $K_{F98}^\prime$ values in Table~\ref{tab_comb4} from the compilation of \citet{fernley98}, and six have $K_{BW}^\prime$ values.  
These data are compared with our $\langle K_s \rangle$ values in the top panel of Figure~\ref{fig_comp}.  
%The top panel shows the $K_{F98}$ values after correction to the 2MASS system.  
The six high quality BW stars have a mean difference $\langle K_s \rangle - K_{BW}^\prime = -0.013$ with a standard deviation of $\sigma = 0.044$ mag.  If we ignore the star with the very large error bar (W~Tuc, for which we  had only one very faint comparison star, leading to the very scattered light curve seen in Figure~\ref{fig_lc3}), we obtain a mean of --0.001 mag with $\sigma = 0.036$ mag.  Most of this scatter can be attributed to the errors in our observations, and the mean suggests that the two sets of photometry are on a consistent system.  
%All six of these stars were observed with SMARTS, providing evidence that these observations are not systematically too bright.

The stars observed by \citet{fernley93b} (reference \#2), after correction to $K_s$,  scatter more widely and have a substantial offset: $\langle K_s \rangle - K_{F98}^\prime = -0.019$ with  $\sigma = 0.054$ mag.  If the bright outlier with the large error bar (V675~Sgr) is omitted, these values become --0.013 and 0.048 mag, respectively.  Part of this scatter can be attributed to errors in our observations.  Subtracting in quadrature our mean $\sigma_K$ of 0.020 mag leaves 0.044 mag attributable to errors in the Fernley magnitudes,   which we list as the magnitude uncertainty $\sigma_{F98}$ in Table~\ref{tab_comb4}. 
%The observations from SMARTS tend to lie below  those from MDM, suggesting they may be systematically brighter.

%Clearly mean magnitude values based on sparse observations without template fitting provide inferior magnitudes.  

The stars with unpublished photometry reported in \citet{fernley98} (reference \#3) also scatter widely, with $\langle K_s \rangle - K_{F98}^\prime = -0.006$ and  $\sigma = 0.121$ mag.  If we omit the two extreme outliers (TY~Aps and AU~Vir), these values become --0.005 and 0.052 mag, respectively, giving no evidence of a systematic offset.   Subtracting our errors in quadrature indicates that 0.048 mag of the scatter can be attributed to errors in the unpublished photometry,   which we list in Table~\ref{tab_comb4} as $\sigma_{F98}$ for these stars. 
  %There is again weak evidence that our SMARTS observations are systematically brighter than our MDM data.
% so we infer that all the unpublished stars behave this way; we also assign these stars an uncertainty of 0.05 mag in Table 4.

%The top panel shows the $K_{F98}$ values as published without adjustment.  The three recalibrated stars \citep{monson17} are in very good agreement with our  $\langle K_s \rangle$ values; their average $\langle K_s \rangle - K_{F98} = 0.023$ with a standard deviation of $\sigma = 0.015$ mag.  
%%suggesting that our SMARTS data are reliably on the 2MASS scale.  
%The six stars with ``ref=1'' do something.  In the bottom panel, we show these stars after correction from their native photometric systems (SAAO for XX, and ? for Cacc) to the 2MASS system using Eqns.~N-M of  \citep{monson17}.  These stars average NN mag fainter? than our data, with $\sigma = MM$ mag, again suggesting that our (MDM/SMARTS?) data are reliably on the 2MASS scale. 

%The stars which were observed less frequently, ref=2 and ref=3, scatter more broadly in both panels.  After correction (the \citet{fernley93b} ``ref=2'' data are on the CTIO system; we assume the unpublished ``ref=3'' data cited in \citet{fernley98} are as well) have mean sig stuff.

%Uncertainties in the period, either due to observational error or actual changes in the star's pulsation \citep{smith95} add scatter to the resulting light curve.

% 2019mar22: original intro PP about Dambis 09,13 was moved to end of Sec.7.1

The lower panel of Figure~\ref{fig_comp} compares our magnitudes with the those of \citet{dambis13}. 
%who added a dozen stars to the set of magnitudes from \citet{dambis09}.  Following a method outlined by \citet{feast08}, these authors phased each star's single-epoch photometry from 2MASS using an external ephemeris and fit a template at fixed phase and amplitude to obtain the intensity-mean magnitude, $\langle K_{DF} \rangle$.  The estimated uncertainty in a typical fit was 0.03 mag, and they reported that most stars had uncertainties in their 2MASS photometry of $\sim$0.02 mag, giving a combined uncertainty of 0.03-0.04 mag.  
The mean difference is $\langle K_s \rangle - K_{D13} = -0.017$ with  $\sigma = 0.061$ mag for the 74 stars in common, all of which had been phased and template-fit by \citet{dambis09}.  
%\citet{dambis09} cautioned that $\sim$8\% of their stars did not have reliable ephemerides, so they did not phase and fit templates to these 2MASS data, and simply adopted the 2MASS magnitude at unknown phase. 
The four outliers (from left to right: V~Ind, V675~Sgr, RV~Cap, and BB~Eri) have lower quality photometry from our SMARTS observations, but none is so different from the other SMARTS results as to create the offsets seen in Figure~\ref{fig_comp}, and so we hypothesize that the errors for these four stars are in the \citet{dambis09} photometry.  If these points are rejected, the mean and standard deviation become --0.022 and 0.046 mag, respectively.  
%MDM/SMARTS offset?
% Dambis did not report which stars suffered this effect.   Reject 4 OL --> mean=-0.023 sigma=0.046

To test the hypothesis that the outliers were caused by the Dambis photometry, we matched the Dambis data with the corrected photometry for all the stars with densely-populated light curves from BW analyses, $K_{BW}^\prime$.
%, corrected their magnitudes to the 2MASS photometric system, and matched them with the \citet{dambis13} photometry.  
Again, all stars in the comparison were phased and fit by \citet{dambis09}, and again there are clear outliers in the lower panel of Figure~\ref{fig_comp} (from left to right: AR~Per, DX~Del, and UU~Vir). Each of these stars was observed as part of a different BW study, making it more likely that the discrepancies resulted from the Dambis photometry.
%difficult to believe that the offsets are due to the BW photometry. 
After rejecting these three outliers, we found the remaining stars had $K_{BW}^\prime - K_{D13}  = -0.028$ with  $\sigma = 0.030$ mag.  The scatter is in agreement with the size of the errors indicated by \citet{dambis09} and \citet{dambis13}, while the mean is consistent with the offset we found between our observations and the Dambis data; it hints that the Dambis data may be systematically faint by $\sim$0.025 mag, though the cause of such an offset is not obvious.

We can also use the information in Figure~\ref{fig_comp} to check whether there is a systematic offset between our data taken at the SMARTS and MDM observatories.  In the top panel, the observations from SMARTS tend to lie below those from MDM when compared with stars from \citet{fernley93b} (reference 2) and with unpublished stars from \citet{fernley98} (reference 3), suggesting they may be systematically brighter.  The offsets are $\langle K_s \rangle - K_{F98}^\prime = -0.041 \pm 0.016$ mag and $+0.001 \pm 0.012$ mag for the reference 2 data compared with data from SMARTS and MDM, respectively.  They are $-0.039 \pm 0.031$ and $+0.023 \pm 0.011$ mag, respectively, for the reference 3 data.  For the comparison shown in the bottom panel of Figure~\ref{fig_comp}, the offsets are $\langle K_s \rangle - K_{D13} = -0.030 \pm 0.007$ and $-0.013 \pm 0.008$ mag for the data from SMARTS and MDM, respectively.  These all suggest that our SMARTS data are systematically brighter than our MDM data.  The single but very significant exception is the comparison in the top panel between our data and the highest quality data, which were obtained from previous BW studies, for which $\langle K_s \rangle - K_{BW}^\prime = -0.001 \pm 0.016$ mag for the SMARTS data (no MDM data were obtained of BW stars).  The weighted mean of the three offsets using the MDM data is $0.000 \pm 0.006$ mag, indicating the MDM data are securely on the 2MASS system, while the weighted mean of the four offsets using the SMARTS data is $-0.028 \pm 0.006$ mag, supporting the idea that these data are systematically too bright.  In calculating each of these offsets, the outlier points discussed above have been removed.  The choices of which stars to consider as outliers and which weighting scheme to utilize have small but meaningful effects on the size of the systematic photometric offset of the SMARTS data, which we estimate to be 0.02-0.03 mag.  The origin of this offset is not known, and we recommend that independent photometric observations be acquired for some of our SMARTS stars to confirm and quantify the size of any offset.

% ref=2
% us_them2s.dat   SM:   -0.060 pm 0.016  7 stars (V675Sgr outlier removed)
% us_them2m.dat   MDM: -0.018 pm 0.012 14 stars
%  Diff = 0.052 pm 0.020 (2.6 sigma)
% ref=3
% us_them3s.dat   SM:   -0.039 pm 0.031  4 stars (2outliers removed)
% us_them3m.dat   MDM: 0.023 pm 0.011  5 stars
%  Diff = 0.062 pm 0.032 (1.9 sigma)
% Dambis09:
%    SM:  -0.032 pm 0.008  38 stars (4 outliers removed)
%   MDM: -0.013 pm 0.008  33 stars
%  Diff =  0.019 pm 0.011 (1.7 sigma)
% ref=1/4
%      SM = -0.001 pm 0.016  5 stars (1 outlier removed, W Tuc)

% Each of the three outliers is from a different group who did the BW analyses.

We conclude that %literature 
photometry which is based on small numbers of photometric observations -- specifically, \citet{dambis09}, the unpublished photometry reported in \citet{fernley98}, and to a lesser extent \citet{fernley93b} -- contains a small fraction of stars with random photometric errors of $\sim$0.1 mag which can not be identified \textit{a priori}.  
%points that were obtained at phases far from the mean.  
This is true even of the \citet{dambis13} photometry that was phased with ephemerides from the literature and fit with a light curve template to obtain a star's intensity-mean magnitude; seven of the 104 stars (7\%) in the comparisons above were outliers.  %It is possible that one or more of the sources of ephemerides used by \citet{dambis09} was not successful in correctly phasing the 2MASS photometry.  
This is in addition to the 32 of 403 stars (8\%) in \citet{dambis13} which had no reliable ephemeris in the literature and were expected to have random offsets of $\sim$0.1 mag.  We found the outlier rates in the \citet{fernley98} $K$-magnitudes to be 5\% for stars with reference~\#2 and 18\% for stars with reference~\#3.  % ( \citet{fsb93} and unpublished, respectively).

% CONCLUDE: ~15% of stars in \citet{dambis13} are outliers; half due to not being phased/fit, and half due to unknown cause in DF phase/fit procedure (maybe one or more ephemerides were no good)
 % CONCLUDE: zero-point uncertainties persist at the ~0.02 mag level in the different sources available in the literature; averaging across several studies will tend to dampen their effect on the resulting absolute mags.  In particular, we provide evidence suggesting that the set of photometry in \citet{dambis13} is ~0.025 mag fainter than the system defined by our MDM data and the BW observations (after correction to the 2MASS system).

%We also note that in some, but not all of the comparisons considered above, our SMARTS observations tend to be $\sim$0.02 mag brighter than the photometry they are compared with, suggesting the presence of a systematic offset of unknown origin.  However, we are reluctant to make an \textit{ad hoc} correction and so proceed with the magnitudes as listed in Table~\ref{tab_finK}.

\subsection{Photometric Means} \label{ssec_wmean}

In an effort to ameliorate the effect of any zero--point offsets, to identify and reject outliers, and to bring as much independent information as possible to bear on determining each star's intensity-mean apparent magnitude, we developed the following scheme for combining the photometric data.  First, we corrected any magnitudes from the literature to the 2MASS photometric system using the methods previously described, and compiled all the available magnitudes for each star   as shown in Table~\ref{tab_comb4}.   We computed a weighted mean for each star using the following weights, $w_i$, and tracked which sources were utilized using a code composed of four integers, $c_i$.  For the stars with densely-populated BW light curves, we adopt $w_1 = 0.009^{-2}$ mag \citep{monson17} and set the first digit of the code to $c_1 = 1$   for the twenty stars from Table~5 of \citet{monson17} including RR~Lyrae itself, and the code  $c_1 = 2$ for the ten BW stars not in \citet{monson17} which we transformed to the 2MASS system ourselves (the prime symbol in Table~\ref{tab_comb4} indicates the transformation). 

For the   stars with our $\langle K_s \rangle$ values  in Table~\ref{tab_comb4} we adopt $w_2 = \sigma_K^{-2}$ and set the second digit $c_2$ to ``1'' if the star was observed with SMARTS or ``2'' if it was observed using MDM.  Code $c_2 = 3$ indicates that both observatories were used to observe YZ~Cap and we hereafter use the weighted mean of $\langle K_s \rangle = 10.355 \pm 0.011$ mag for this star; while $c_2 = 4$ indicates the direct 2MASS observation of NSV~660. We also increased the $\langle K_s \rangle$ value for each star observed with SMARTS by 0.028 mag to account for the offset estimated above.  

The stars in the compilation of \citet{fernley98}, shown in Table~\ref{tab_comb4} under the heading $K^\prime_{F98}$, are given $w_3 = 0.044^{-2}$ and $c_3 = 2$ in the third digit if it was from \citet{fernley93b} (60 stars,   reference \#2 in Table~\ref{tab_targets}), or $w_3 = 0.048^{-2}$ and $c_3 = 3$ if the data was unpublished (20 stars,   reference \#3).  Finally, each of the stars in the compilation (except NSV~660) has a magnitude and uncertainty, $\sigma_{D13}$, from \citet{dambis13}, which we weight with $w_4 = \sigma_{D13}^{-2}$ mag and identify with $c_4 = 1$ in the fourth digit.  Any star for which data was not available will have a value of zero in the corresponding digit of the code.  For example, a star having a code of ``1021'' indicates the star has photometry from a high density BW source, no observation from Table~\ref{tab_finK}, and had photometric values in \citet{fernley93b} and  \citet{dambis13}.    The combined digits of this code for each star are listed in Table~\ref{tab_comb4} under the column labeled ``Code$_K$.'' 

So that we could detect and reject outlier photometry for a given star, we retained only stars with two or more photometric values and computed the weighted mean and the difference of each magnitude from that mean.  Data from any stars with differences %($| \Delta K |$) 
more than $\pm0.1$ mag from the weighted mean were inspected visually and if one source was clearly discrepant, it was rejected and identified with a ``9'' in the digit of the code corresponding to that photometric source (four stars).  For seven stars with only two observations from lower-quality sources, it was difficult to distinguish outliers, and we retained the potentially discrepant photometry.  Two other stars are noteworthy: V675~Sgr and RV~Cap both had three measures of lower quality, and for both stars, our $\langle K_s \rangle$ value was bracketed by discrepant values ($| \Delta K | > 0.1$ mag) from \citet{fernley98} and \citet{dambis13}; since no outlier could be identified, we retained all three values when computing the weighted mean magnitude for each star.  Of the 13 stars with photometric deviations larger than 0.1 mag, only RV~Cap is known to exhibit the Blazhko effect, so in general these cyclic variations in amplitude can not be responsible for the outlier behavior seen in Figure~\ref{fig_comp}.

%[TO DO?  Move the mean K, warning flag, mean Av, and reference code from Table~\ref{tab_gaia} to Table~\ref{tab_finK}, and report only the $K_0$ in Table~\ref{tab_gaia}. ]
For each of the remaining 146 stars, the weighted mean magnitude $\bar{\langle K \rangle}$ and its error $\sigma_{\bar{K}}$ are shown in Table~\ref{tab_comb4}.   In the column labeled ``Flag'' we identify   stars whose data scatter more broadly than suggested by the weighted error; these stars are strong candidates for attention in future photometric programs.   Overall, our weighting scheme places high weight on the BW and MDM photometry, while weighting SMARTS photometry and that from the other sources at similar, lower levels.

\subsection{Reddening Means} \label{ssec_meanak}

We used an analogous scheme for weighting and identifying the contributions of the three interstellar reddening values listed in Table~\ref{tab_targets}.  %Stars with low reddening, $E_{SD11} < 0.1$ mag receive double weight on the results of the dust maps... 
We used weights $w_i = \sigma_i^{-2}$, where $\sigma_2 = 0.03$ mag for the \citet{blanco92} reddening $E(B-V)_{B92}$, and $\sigma_3 = 0.03$ mag for the \citet{fernley98} reddening $E(B-V)_{F98}$, as discussed in Sec.~\ref{sec_sel}.
%the online Appendix.
%two RRL color-based estimates, and the the codes $c_{2,3} = 1$ if data was available for the star, or 0 if it was not.  
For the dust-based reddening estimate, we adopted the uncertainty estimate from \citet{sfd98} of $\sigma_1 = 0.16~E(B-V)_{SD11}$, but set a minimum value of 0.01 mag so it would not unduly dominate the weighting when the reddening was small.  We did not include these reddenings in the weighted mean when $E(B-V)_{SD11} > 0.20$ mag, because they are systematically overestimated as shown in Sec.~\ref{sec_sel}, and flagged them   with $a_1 = 9$ in the first digit of a three-integer identification code.   Otherwise, $a_i = 1$ indicates  that a value was used in the weighted mean, while 0 indicates the value was not available.  Using $A_K = 0.36~E(B-V)$ from \citet{cardelli89}, we converted the weighted mean reddening and its error into interstellar absorption, $A_K$, and its error $\sigma_A$, which we list in Table~\ref{tab_comb4} for each star along with the integer code labeled ``Code$_A$.''

%New v2
We note that \citet{dambis13} provided an independent set of extinction values, $A_V$, in their compilation, obtained from an iterative solution to the 3D dust distribution model of \citet{drimmel03}.  We again used $A_K = 0.36~E(B-V)$ along with $A_V = 3.1~E(B-V)$ from \citet{cardelli89}  to convert these optical extinction values to infrared, and compare them with our own.  For the 145 stars in common, the mean difference $A_K - A_K(D13) = -0.0002$ mag with a standard deviation $\sigma = 0.007$ mag, indicating outstanding agreement between our data sets.

%\bibitem[] {drimmel03} Drimmel, R., Cabrera-Lavers, A., \& L\'opez-Corredoira, M. 2003, \aap, 409, 205
% A large-scale three-dimensional model of Galactic extinction is presented based on the Galactic dust distribution model 

%Things to try: (1) make MDM points solid, SMARTS open and see whether offset-sigma differ, comment if SMARTS is brighter like YZ Cap; (2) ID the 4-5 outliers in Dambis; my bad?; (3) match Dambis with Fernly98 and compare to see if the scatter gets larger (they have bigger errors) or smaller (I do) make ref1 points special since carry more weight; if the their problem, average Dambis+F98 of stars at bottom of Table1 for use with parallaxes in Table 4.

%Using the lessons learned from this analysis, we compile a final list of photometry in Table~\ref{tab_gaia}.  Our $\langle K_s \rangle$ and $\sigma_K$ values for the 75? RRL in Table~\ref{tab_finK} are listed first, with the reference code 0 (differentiate MDM/SMARTS?).  Stars with densely-populated light curves, after correction to the 2MASS photometric system, are listed with the reference code 4; for each these stars we adopt an uncertainty of 0.009 mag following \citet{monson17}.  

% NSV 660 = ?  our analysis of 2MASS time-series photom.
% FSB93 bad 110 + dambis? 

% Table~\label{tab_gaia}
\begin{deluxetable*}{lcrrrrrrrrrr}
%aastex52 \tabletypesize{\scriptsize}
%aastex52 \rotate
%\tablecaption{Parallax and Absolute Magnitudes \label{tab_gaia}}
\tablecaption{Final Data for Absolute Magnitude Solutions \label{tab_gaia}}
% table7_sub2.txt --> tab7b.txt
%aastex52 \tablewidth{0pt}
\tablehead{
\colhead{Star}  & \colhead{Type} & \colhead{$m_{K,o}$} & \colhead{$\sigma_{K,o}$} & 
\colhead{ [Fe/H] } &  \colhead{$\sigma_\mathrm{ [Fe/H] }$} & \colhead{$\log{P_f}$} &  \colhead{Gaia ID}  & 
\colhead{$\varpi$} & \colhead{$\sigma_\varpi$} &
\colhead{$M_{K_s}$} & \colhead{$\sigma_{M_K}$}   
}
%\colnumbers
\startdata
SW~And   & ab & 8.494 & 0.009 & --0.24 & 0.09 & --0.354304 & 2857456207478683776 & 1.7797 & 0.1636 & --0.254 & 0.210 \\
XX~And   & ab & 9.429 & 0.027 & --1.94 & 0.16 & --0.141017 & 370067649378653440 & 0.6950 & 0.0463 & --1.361 & 0.152 \\
AT~And   & ab & 9.026 & 0.028 & --1.18 & 0.13 & --0.209776 & 1925406252226143104 & 2.1779 & 0.2715 & +0.716 & 0.291 \\
DM~And  & ab & 10.515 & 0.009 & --2.32 & 0.15 & --0.200391 & 1912453760434108928 & 0.6000 & 0.0617 & --0.594 & 0.236 \\
\nodata & \nodata & \nodata & \nodata & \nodata & \nodata & \nodata & \nodata  & \nodata & \nodata  & \nodata  & \nodata  \\
AV~Vir  & ab & 10.547 & 0.013 & --1.25 & 0.16 & --0.182496 & 3731723090075245696 & 0.5297 & 0.0470 & --0.833 & 0.202 \\
BN~Vul  & ab & 8.557 & 0.027 & --1.61 & 0.22 & --0.226122 & 2022835523801236864 & 1.4014 & 0.0301 & --0.710 & 0.054 \\
NSV~660 & ab & 13.976 & 0.020 & --1.31 & 0.10 & --0.196141 & 2507784713545431424 & 0.0567 & 0.0444 & --2.228 & 3.318 \\
\enddata
%\tablenotetext{a}{The code identifying the sources of the photometry in the weighted mean $K$ magnitude (see Sec.~\ref{sec_compare}).  }
\tablecomments{(This table is available in its entirety in machine-readable form.)}
%\tablecomments{Table \ref{tab_gaia} is published in its entirety in the electronic edition of the \textit{Astronomical Journal}. A portion is shown here for guidance regarding its form and content.}
\end{deluxetable*}

%Talk about using the $K_S$ mags from 2MASS of the variable and check star to confirm the calibration.  Footnote that LC is available for each star at ACL webpage.

\bigskip
\bigskip
\bigskip
\bigskip
\bigskip

\section{{\textit Gaia} Parallaxes and the RR~Lyrae Luminosity Calibration}  \label{sec_gaia}

%NB: 2019may23 -- completely removed Sec.8 and replaced with Brian's stuff datad 2019 Apr 5.

  In Table~\ref{tab_gaia}, we compile the data needed to perform period-luminosity-metallicity (PLZ) fits on our RRL sample.  The column labeled $m_{K,o}$ contains the mean apparent magnitude $\bar{\langle K \rangle}$ from Table~\ref{tab_comb4} after correction for interstellar absorption $A_K$, and the column labeled $\sigma_{K,o}$ lists its uncertainty computed by adding the uncertainties from those values in quadrature.  The metallicity [Fe/H] was taken from Table~\ref{tab_targets}, using the value from \citet{fernley98} if one is available, and using the value from \citet{layden94} if it was not.  The weights $\sigma_\mathrm{ [Fe/H] }$ are as described in Sec.~\ref{ssec_samz}.  The logarithm of the adopted period from Table~\ref{tab_finK} is presented as $\log{P_f}$, where any star with a pulsation type ``c'' has been fundametalized by adding 0.127 to the logarithm of its period.  

We searched the \textit{Gaia} Data Release 2 (DR2) catalog\footnote{Accessed 2018 May 11 via the Vizier service at \url{http://vizier.u-strasbg.fr/viz-bin/VizieR}.} \citep{gaiaDR1,gaiaDR2cat} using a search cone with 2-4 arcsec radius around the 2MASS coordinates listed in Table~\ref{tab_targets}.  In every case   we obtained only one match, and we checked this star's parallax, proper motion, and magnitude with values from the literature to ensure that we had correclty identified the RRL.   Each star's unique \textit{Gaia} identifier number is listed in Table~\ref{tab_gaia} along with its parallax value $\varpi$ and  its uncertainty $\sigma_{\varpi}$ in units of milli-arcseconds (mas).  
The parallax for the star RR~Lyrae is negative as its mean $G$ magnitude was determined incorrectly 
%%(10 mag too faint)
in DR2 \citep{muraveva18},   and so we do not use this star in our analysis.
%report the value from the \textit{Gaia} DR1 in Table~\ref{tab_gaia}. } 

%we correctly identified the RRL.  
  The DR2 catalog is known to contain spurious parallaxes which formally have small parallax uncertainties. Spurious  parallaxes typically have a large astrometric $\chi^2$ and the \textit{Gaia} team recommend that one calculates the Unit Weight Error \citep{lindegren18}
%(Lindegren et. al.\ 2018)\footnote{\textbf{Presented at the IAU GA Division A meeting on 27 August 2018, and available online at  https://www.cosmos.esa.int/web/gaia/dr2-known-issues .}} 
to find spurious astrometric solutions 
$$
\mathrm{UWE}  = \sqrt{ \chi^2/(N - 1) }
$$
where $\chi^2 = $ \texttt{astrometric\_chi2\_al} and\hfill\\ $N = $ \texttt{astrometric\_n\_good\_obs\_al} are available in the DR2 catalog.  Large values of UWE indicate a bad astrometric solution.  \citet{lindegren18} further recommend that the RUWE, which depends on the color and magnitude is a better way to determine if an astrometric solution is spurious.  However, since RRL have variable colors and magnitudes, it is not possible to calculate RUWE from the data provided in DR2, and so we elect to use the UWE to search for spurious parallaxes.  The UWE was calculated for each of our stars, and two %three
stars were found to have large UWE values $>2.5$: VV~Peg (UWE $ = 6.9$) and AT~And  (UWE = $12.8$).  % and RR~Lyr (UWE $= 24.3$).  
We do not use these %three 
stars in any further analysis, leaving our RRL sample with 143 stars.  
%The parallax values and their uncertainties} are listed in Table~\ref{tab_gaia} as $\varpi$ and $\sigma_{\varpi}$ in units of mas.  The parallax for the star RR~Lyrae  is negative as its mean $G$ magnitude was determined incorrectly %(10 mag too faint)
%in DR2 \citep{muraveva18},  \textbf{and the  value from DR1 is reported in Table~\ref{tab_gaia}.

The uncertainties in the DR2 parallaxes are known to be underestimated by $\sim$30$\%$ \citep{gaiaDR2use,gaiaDR2valid} for the brighter sources.    \citet{lindegren18} have presented a tentative calibration of the true, external error in the \textit{Gaia} DR2 parallaxes, $\sigma_{ext} = \sqrt{1.08^2 \sigma_i^2 + \sigma_s^2}$, where $\sigma_i$ is the internal parallax error reported for each star in DR2, and $\sigma_s = 0.021\,$mas for bright ($G \la 13$) stars and $\sigma_s = 0.043\,$mas for faint ($G > 13$) stars.  We have applied this correction to the reported parallax uncertainties in all of our analysis,
 though we list the DR2 values as provided in Table~\ref{tab_gaia}.    
 Using these increased uncertainties, the median value of $\varpi/\sigma_\varpi = 18.2$ for stars which we use in our analysis,  with only  two stars  having $\varpi/\sigma_\varpi < 5.0$, indicating the high quality of the parallaxes now available.

\begin{figure}
%aastex52 \epsscale{.80}
%\plotone{periodMk.pdf}
%\plotone{periodMkbad.pdf}
%\plotone{fig10b.pdf}
\plotone{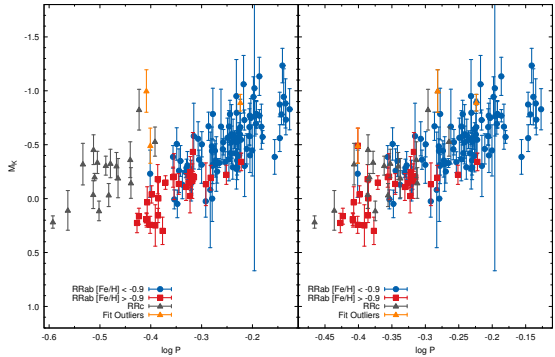}
\caption{The parallax-based absolute magnitude $M_{K_s}$ from Table~\ref{tab_gaia} is plotted against the logarithm of the period for 143 RRL stars.  The RRc stars had their periods fundamentalized by adding 0.127 to $\log P$ in the right panel.  The RRab stars are divided into two metallicity groups as indicated in the legend,  corresponding to the disk and halo kinematic populations (see Sec.~\ref{ssec_samchar});  there are no metal-rich RRc stars in this sample.   Three stars that are outliers from our PLZ fit (see Sec.~\ref{ssec_plzobs}) are marked in orange.   A  DR2 parallax zero-point error of $-0.036\,$mas was assumed when determining $M_{K_s}$.  
    \label{fig_Mk1}}
\end{figure}

For each star in Table~\ref{tab_gaia} we calculated the absolute magnitude (along with its uncertainty, $\sigma_{M_K}$) in the 2MASS $K_s$ photometric system using
\begin{equation}
%  M_{K_s}  = \bar{\langle K \rangle} - A_K +  5~\rm{log}_{10} (\varpi) - 10,
  M_{K_s}  = m_{K,o} +  5~\rm{log}_{10} (\varpi) - 10.
  \label{eqnMk}
\end{equation}
%where $\varpi$ is the parallax in milli-arcseconds.     
These values are listed in Table~\ref{tab_gaia} and plotted against the logarithm of the stars' pulsation periods in Figure~\ref{fig_Mk1} for all 143 RRL  with reliable DR2 parallaxes in our sample.  We note that \cite{gaiaDR2use} strongly recommend against using the parallax to directly infer the distance (as in the above equation) when deriving astrophysical parameters. Given that our sample selection was not based upon DR2 data, and the generally high quality of the DR2 parallaxes in our sample, much of the reasoning presented by \cite{gaiaDR2use} does not apply to our case.    Figure \ref{fig_Mk1} is informative, though we don't use it our subsequent analysis. 
  
 %There appears to be one outlier in Figure~\ref{fig_Mk1}: the star AT~And is anomalously faint at $M_{K_s} = +0.72 \pm 0.36$ mag, which is $\sim 1.4\,$mag fainter than the bulk of the stars at its period.  
 %However, when we fit for the PLZ relation (see below) AT~And observed data differs from its fitted data by $2.8\,\sigma$ and so is not a significant outlier. 
The c-type RRL  form a grouping on the left with shorter periods.  A  trend with log~$P$ is apparent for both RRab and RRc stars.  We note that if the period of the RRc stars is ``fundamentalized'' by adding 0.127 to $\log P$, then the RRc stars appear well mixed with the RRab stars.  There is a clear tendency for the metal-rich RRab stars to be fainter than the metal-poor RRab stars, indicating that a 
%period-luminoisty-metallicity (PLZ) 
PLZ relation exists in the data.

% New subsection
\subsection{PLZ Fitting Procedure} \label{ssec_plzfit}

To explore the PLZ relation for RRLyr stars in the $K_s$ band we use the astrometric based luminosity (ABL) prescription of \cite{ABL}.  The ABL approach avoids  biases caused by converting to magnitudes, allows the use of low-quality parallax measurements (including NSV 660) and has been shown to yield similar results to a Baysian analysis when examining PL and PLZ relations \citep{gaiaDR1}.   Because we use the data for all the stars in our sample and do not truncate the sample at an ``observed'' relative error, $\varpi/\sigma_\varpi$, we do not introduce a systematic error in absolute magnitude of the type discussed by \citet{lutzkelker73}, a point made by \cite{ABL} along with some other noteworthy interpretations. 
\cite{ABL} defined the ABL to be
\begin{equation}
a_K = 10^{0.2M_{K}} = \varpi 10^{0.2m_{K,o} - 2}
\label{eqnabl}
\end{equation}
where $M_K$ is the $K_s$ absolute magnitude, $\varpi$ is the parallax in mas and $m_{K,o}$ is the extinction corrected apparent magnitude.  The  PLZ relation may be written as 
\begin{equation} 
M_K  = \beta_1 +  \beta_2 (\log P +0.27) + \beta_3 ( [\mathrm{Fe/H}] + 1.3) \, , \\
\end{equation}
where one determines the coefficients $\beta$ from a fit to the data.  The addition of 0.28 to $\log P$ and 1.3 to [Fe/H] minimizes the uncertainty in the zero-point found by the fit, as the average value of $\log P = -0.27$ and the average $[\mathrm{Fe/H}]  = -1.3$ in this  dataset. 
In the ABL approach, the following equation is used to determine the PLZ relation 
\begin{equation}
 \varpi 10^{0.2m_{K,o} - 2} = 10^{0.2 [\beta_1 +  \beta_2 (\log P + 0.27) + \beta_3 ([\mathrm{Fe/H}] + 1.3) ] }  \; .
 \label{eqnABLfit}
\end{equation}

In analyzing the global luminosity properties of DR2 stars, one must take into account that there exists a global zero-point parallax error in DR2 \citep{gaiaDR2cat,gaiaDR2valid,gaiaDR2use}.  An analysis of quasars indicates the zero-point error is  $-0.029\,$mas for fainter objects \citep{gaiaDR2cat}, and there are indications from the quasar sample that the zero-point error increases ($\sim$0.05~mas) at  brighter magnitudes \citep{gaiaDR2cat}.  All of the quasars are fainter than our RR Lyr dataset.  External comparisons to relatively bright stars with VLBI or \textit{HST} FGS parallaxes indicate a zero-point error of $-0.07\pm 0.03\,$mas and $\-0.01\pm 0.02\,$mas respectively \citep{gaiaDR2valid}. These comparisons, along with other analysis led the {\textit Gaia} collaboration to conclude that the global zero-point is less than $0.1\,$mas.   This has been verified by \cite{stassun18}, who compare {\textit Gaia} DR2 distances to distances determined to relatively bright eclipsing binaries and find a DR2 zero-point error of $-0.08\pm 0.03\,$mas. 
 The parallax zero-point error varies spatially \citep{gaiaDR2cat,gaiaDR2valid}, but since our RR Lyr stars are randomly distributed across the sky, it is only the global zero-point error which is of importance to our PLZ analysis.  Allowing for a global zero-point error requires the use of an implicit fitting function,   
 \begin{eqnarray}
f  & \equiv & 10^{0.2 [\beta_1 +  \beta_2 (\log P + 0.27) + \beta_3 ([\mathrm{Fe/H}] + 1.3) ] } \nonumber \\
 & -&   (\varpi + \beta_4) 10^{0.2m_{K,o} - 2} = 0
 \label{eqnfitfunction}
\end{eqnarray}
where $\beta_4 = \pi_{zp}$ is determined as part of the fitting process.

In performing fits to data, one weights the fit by the uncertainty in the data.  Many of the stars in our dataset have very well determined parallaxes and mean magnitudes, with very small formal uncertainties.  However, the recent paper by \cite{braga18} which presents $JHK$ photometry of RRL in $\omega\,$Centauri suggests that there is an intrinsic dispersion in the PLZ relation.  \cite{braga18}  fit the PLZ relation for a large number of stars spanning $-2.3 < [\mathrm{Fe/H}] < -1.3$.  Depending on which metallicities are employed \citep{rey00,sollima06,braga16},  standard deviations of the RRL around the best-fit PLZ relations are 0.045, 0.029, and 0.039 mag, respectively for their  ``global'' solutions  where RRab and fundamentalized RRc are in the same fit.  The residual magnitude distributions in their Figures~19, 20 and 21 appear to be roughly Gaussian.
%, with no obvious tail to brighter mags (suggesting that evolution of BHB stars across the instability strip at brighter magnitudes was accounted  for by the period term of the fit). 
These standard deviations  are  larger than the photometric uncertainties in their data as the the median of their $K_s$ uncertainties is 0.008 mag for 198 RRL in their Table~2.  Taking this at face value and subtracting it in quadrature from the above standard deviations gives an estimate of the intrinsic dispersion in $M_K$ in the PLZ relation of approximately 0.03 to 0.04 mag. To take into account the intrinsic dispersion in the PLZ relation, when performing the fits, the uncertainty in the $K_s$ magnitude is found by adding the photometric uncertainty in quadrature with an intrinsic dispersion of $\sigma_{M_K, \, \mathrm{intrinsic} } = 0.04\,$mag.

% New subsection
\subsection{Monte Carlo Tests} \label{ssec_mctest}

The implicit nonlinear weighted orthogonal distance fitting was performed using ODRPACK95 \citep{odrpack95}, which is an updated version of ODRPACK \citep{odrpack}.  This fitting routine takes into account the uncertainties in the photometry, extinction corrections, parallaxes, [Fe/H] values and periods.    To test the reliability of the fitting procedure and the reported uncertainties, we conducted a Monte Carlo simulation with synthetic data.  In this simulation, we randomly created data sets (with $N= 143$ stars) whose distribution in apparent $K_s$ magnitude, period and [Fe/H], along with their associated uncertainties, were all drawn randomly from the observed distributions.  A PLZ relation (including an intrinsic $M_K$ dispersion of 0.04 mag) was specified and used to determine a `true' parallax for the star.  This true parallax was then used to create a simulated parallax by randomly selecting a value from a Gaussian distribution whose mean is the `true' parallax, and whose standard deviation was drawn randomly from the distribution of standard deviations in the  actual data.   A specified zero-point error was then added to each simulated parallax.  This results in a simulated dataset whose properties resemble the actual dataset.  The simulation was performed 1000 times for a number of different choices of the PLZ relation and parallax zero-point error.  The simulated data was then used by our fitting program to calculate the PLZ relation.

These initial tests indicated that ODRPACK95 was reliably determining the zero-points in our fitting function ($\beta_1$ and $\beta_4$ in equation~\ref{eqnfitfunction}), but did not determine the slopes ($\beta_2$ and $\beta_3$ in equation~\ref{eqnfitfunction})  reliably.  As a result, we adopted a two-step procedure for our fits.  After performing the implicit fit with ODRPACK95 and determining the DR2 global parallax offset, we then performed an explicit nonlinear fit (equation~\ref{eqnABLfit}) using \texttt{gnuplot}\footnote{http://www.gnuplot.info;  see http://www.gnuplot.info/docs\_5.0/gnuplot.pdf for a discussion of the nonlinear fit routine}.  The Monte Carlo tests indicate that this two-step  fitting procedure does an excellent job of recovering the input PLZ relation and the  parallax zero-point, including estimating the uncertainties in the coefficients.  The latter point is demonstrated by the fact that the standard deviation of the Monte Carlo results ($\sigma_\mathrm{MC}$) is very similar to the average (over all the Monte Carlo simulations) of the errors reported by the fitting procedure ($\langle \mathrm{MC~Err} \rangle$).
%($<\mathrm{MC~Err}>$). 
The results of our Monte Carlo tests for our two-step  fitting procedure are summarized in Table~\ref{tab_MC}.  The first two Monte Carlo tests shown in Table~\ref{tab_MC} use identical input parameters, and the differences in the output reflect the statistical uncertainty in our Monte Carlo.

%% This is where Table 8 MC test of ABL was originally

In performing the nonlinear fitting with the actual data,  the goodness of fit was determined by an analysis of the normalized fit residuals ($R_i \equiv f_i/\sigma_{f_{i}}$, where $i$ represents the $i^\mathrm{th}$ star in the observed dataset), as $\chi^2$ analysis is not appropriate for a nonlinear fit \citep{andrae}. The fit residuals should follow a Gaussian distribution, with a mean $\mu = 0$ and variance $\sigma^2 = 1$.   

Since our fitting function is a fairly complicated nonlinear function, uncertainty in its value for a given star was determined through a Monte Carlo simulation, where specific values for each of the variables  ($\beta_1; \beta_2; \beta_3; \beta_4; \log P;  [\mathrm{Fe/H]}; m_{K,o}, \varpi $) was randomly drawn from a Gaussian distribution whose mean and standard deviation are given by their tabulated values for the observed values  or returned by the two-step fitting procedure for the fitted parameters.  This was repeated 1000 times for each  star, and the resulting distribution of $f_i$ was used to determine the uncertainty, $\sigma_{f_i}$. 

\startlongtable
\begin{deluxetable}{lccccc}
\tabletypesize{\tiny}
\tablecaption{Monte Carlo Test of ABL Fitting Procedure \label{tab_MC}}
\tablehead{
%& &
%\colhead{MC} & 
%& \colhead{ $-$ Monte}  &
%\colhead{Carlo} & 
%\colhead{Input $-$} \\
\colhead{Parameter} &
\colhead{Input} &
\colhead{MCmean} &
%\colhead{MC Mean} &
\colhead{Difference} &
\colhead{$\sigma_\mathrm{MC}$} &
\colhead{ $\langle \mathrm{MC~Err} \rangle$ }
%\colhead{ $<\mathrm{MC~Err}>$ }
}
\startdata
PLZ zero-point: & $ -0.450 $ & $ -0.447 $ & $ -0.003 $ & $ 0.026 $ & $ 0.028 $ \\
PLZ Period slope: & $ -2.700 $ & $ -2.697 $ & $ -0.003 $ & $ 0.148 $ & $ 0.146 $ \\
PLZ [Fe/H] slope: & $ 0.300 $ & $ 0.300 $ & $ 0.000 $ & $ 0.022 $ & $ 0.022 $ \\
Parallax zero-point: & $ -0.040 $ & $ -0.040 $ & $ 0.000 $ & $ 0.011 $ & $ 0.011 $ \\
 ~~ \\
PLZ zero-point: & $ -0.450 $ & $ -0.446 $ & $ -0.004 $ & $ 0.027 $ & $ 0.028 $ \\
PLZ Period slope: & $ -2.700 $ & $ -2.696 $ & $ -0.004 $ & $ 0.153 $ & $ 0.146 $ \\
PLZ [Fe/H] slope: & $ 0.300 $ & $ 0.300 $ & $ 0.000 $ & $ 0.023 $ & $ 0.022 $ \\
Parallax zero-point: & $ -0.040 $ & $ -0.041 $ & $ 0.001 $ & $ 0.011 $ & $ 0.011 $ \\
~~\\
PLZ zero-point: & $ -0.450 $ & $ -0.447 $ & $ -0.003 $ & $ 0.027 $ & $ 0.028 $ \\
PLZ Period slope: & $ -2.700 $ & $ -2.692 $ & $ -0.008 $ & $ 0.155 $ & $ 0.144 $ \\
PLZ [Fe/H] slope: & $ 0.300 $ & $ 0.300 $ & $ 0.000 $ & $ 0.023 $ & $ 0.022 $ \\
Parallax zero-point: & $ -0.080 $ & $ -0.080 $ & $ 0.000 $ & $ 0.011 $ & $ 0.011 $ \\
~~ \\
PLZ zero-point: & $ -0.450 $ & $ -0.449 $ & $ -0.001 $ & $ 0.024 $ & $ 0.024 $ \\
PLZ Period slope: & $ -2.700 $ & $ -2.694 $ & $ -0.006 $ & $ 0.154 $ & $ 0.145 $ \\
PLZ [Fe/H] slope: & $ 0.100 $ & $ 0.101 $ & $ -0.001 $ & $ 0.023 $ & $ 0.022 $ \\
Parallax zero-point: & $ -0.080 $ & $ -0.080 $ & $ 0.000 $ & $ 0.010 $ & $ 0.010 $ \\
~~\\
PLZ zero-point: & $ -0.450 $ & $ -0.445 $ & $ -0.005 $ & $ 0.025 $ & $ 0.028 $ \\
PLZ Period slope: & $ -2.400 $ & $ -2.395 $ & $ -0.005 $ & $ 0.153 $ & $ 0.146 $ \\
PLZ [Fe/H] slope: & $ 0.300 $ & $ 0.300 $ & $ 0.000 $ & $ 0.022 $ & $ 0.022 $ \\
Parallax zero-point: & $ -0.040 $ & $ -0.041 $ & $ 0.001 $ & $ 0.010 $ & $ 0.011 $ \\
~~\\
PLZ zero-point: & $ -0.550 $ & $ -0.547 $ & $ -0.003 $ & $ 0.025 $ & $ 0.027 $ \\
PLZ Period slope: & $ -2.600 $ & $ -2.589 $ & $ -0.011 $ & $ 0.152 $ & $ 0.153 $ \\
PLZ [Fe/H] slope: & $ 0.200 $ & $ 0.202 $ & $ -0.002 $ & $ 0.024 $ & $ 0.023 $ \\
Parallax zero-point: & $ -0.030 $ & $ -0.030 $ & $ 0.000 $ & $ 0.010 $ & $ 0.011 $ \\
~~ \\
PLZ zero-point: & $ -0.550 $ & $ -0.548 $ & $ -0.002 $ & $ 0.025 $ & $ 0.027 $ \\
PLZ Period slope: & $ -2.600 $ & $ -2.598 $ & $ -0.002 $ & $ 0.160 $ & $ 0.155 $ \\
PLZ [Fe/H] slope: & $ 0.200 $ & $ 0.200 $ & $ 0.000 $ & $ 0.024 $ & $ 0.023 $ \\
Parallax zero-point: & $ -0.010 $ & $ -0.010 $ & $ 0.000 $ & $ 0.010 $ & $ 0.011 $ \\
~~ \\
PLZ zero-point: & $ -0.450 $ & $ -0.448 $ & $ -0.002 $ & $ 0.023 $ & $ 0.024 $ \\
PLZ Period slope: & $ -3.000 $ & $ -2.997 $ & $ -0.003 $ & $ 0.157 $ & $ 0.146 $ \\
PLZ [Fe/H] slope: & $ 0.100 $ & $ 0.100 $ & $ 0.000 $ & $ 0.023 $ & $ 0.022 $ \\
Parallax zero-point: & $ -0.030 $ & $ -0.030 $ & $ 0.000 $ & $ 0.010 $ & $ 0.010 $ \\
~~ \\
PLZ zero-point: & $ -0.450 $ & $ -0.450 $ & $ 0.000 $ & $ 0.024 $ & $ 0.024 $ \\
PLZ Period slope: & $ -2.800 $ & $ -2.800 $ & $ 0.000 $ & $ 0.151 $ & $ 0.147 $ \\
PLZ [Fe/H] slope: & $ 0.050 $ & $ 0.050 $ & $ 0.000 $ & $ 0.022 $ & $ 0.022 $ \\
Parallax zero-point: & $ -0.030 $ & $ -0.029 $ & $ -0.001 $ & $ 0.010 $ & $ 0.010 $ \\
~~\\
PLZ zero-point: &  $-0.500   $&$ -0.496 $&$   -0.004  $&$   0.026   $&$  0.026 $\\
PLZ Period slope: &$ -2.300 $&$   -2.292  $&$  -0.008  $&$   0.159   $&$  0.150$\\
PLZ [Fe/H] slope: & $0.180  $ &$  0.180  $&$  0.000  $&$   0.023   $&$  0.023$\\
Parallax zero-point: &$ -0.050$&$    -0.051  $&$   0.001    $&$ 0.011   $&$  0.011$\\
 \enddata
 \end{deluxetable}

% New subsection
\subsection{PLZ of the Observed RRL} \label{ssec_plzobs}

In order to evaluate how different [Fe/H] scales and weighting schemes may impact our fitted PLZ relations, we used four different datasets in the  analysis of our observed RRL, each with a slightly different way of determining [Fe/H]:

\begin{description}
%\item[F0]   \textbf{No shift is applied to the F98 or L94 [Fe/H] estimates.  A uniform uncertainty of $\pm 0.20\,$dex is assumed for the [Fe/H] values.}
\item[F0]   The [Fe/H] estimates from \citet{fernley98} and \citet{layden94} were adopted without correction, and a uniform uncertainty of $\pm 0.20\,$dex was assumed for each [Fe/H] value.

\item[F1] No correction was applied to the  \citet{fernley98} and \citet{layden94}  [Fe/H] estimates, and the weights $\sigma_\mathrm{ [Fe/H] }$ were adopted from Table~\ref{tab_gaia}.

\item[F2]  The \citet{fernley98} [Fe/H] values were shifted by $-0.06\,$dex to put  them onto the \citet{layden94}  [Fe/H] scale, and the weights were adopted from Table~\ref{tab_gaia}. 

\item[F3] The \citet{layden94}  [Fe/H] values were shifted by $+0.06\,$dex to put them onto the \citet{fernley98}  scale, and the weights were adopted from Table~\ref{tab_gaia}. 
\end{description}

\begin{table*}
%\tabletypesize{\tiny}
%\rotate
\caption{P-L-Z Fits of Observed RRL}
\begin{center}
\begin{tabular}{lcccccc}
\hline\hline
& \multicolumn{3}{c}{PLZ} & & $\chi^2$\\
\cline{2-4}
Dataset & 
Zero-point &
$\log P$ slope & 
[Fe/H] slope  & 
$ \pi_{zp} $ &
residuals\\  
\hline
~~\\[-12pt]
F0 & $ -0.427 \pm  0.035 $&$ -2.87 \pm  0.19 $&$ 0.122 \pm  0.024 $&$ -0.036 \pm  0.015 $& 1.15 \\
F0 no outliers & $ -0.408 \pm  0.032 $&$ -2.82 \pm  0.17 $&$ 0.122 \pm  0.022 $&$ -0.040 \pm  0.013 $& 1.07 \\
F1 & $ -0.424 \pm  0.033 $&$ -2.86 \pm  0.19 $&$ 0.122 \pm  0.024 $&$ -0.037 \pm  0.014 $& 1.16 \\
F1 no outliers & $ -0.405 \pm  0.031 $&$ -2.81 \pm  0.17 $&$ 0.122 \pm  0.022 $&$ -0.042 \pm  0.013 $& 1.08 \\
F2 & $ -0.414 \pm  0.033 $&$ -2.86 \pm  0.19 $&$ 0.123 \pm  0.024 $&$ -0.038 \pm  0.014 $& 1.16 \\
F2 no outliers & $ -0.394 \pm  0.031 $&$ -2.81 \pm  0.17 $&$ 0.123 \pm  0.022 $&$ -0.043 \pm  0.013 $& 1.08 \\
F3 & $ -0.423 \pm  0.033 $&$ -2.86 \pm  0.19 $&$ 0.123 \pm  0.024 $&$ -0.038 \pm  0.014 $& 1.16 \\
F3 no outliers & $ -0.403 \pm  0.031 $&$ -2.81 \pm  0.17 $&$ 0.123 \pm  0.022 $&$ -0.043 \pm  0.013 $& 1.08 \\
\hline
\end{tabular}
\end{center}
\label{tab_fit1}
\end{table*}%

Results of fitting the PLZ relation to the different datasets are given in Table \ref{tab_fit1}.  The last column in the table gives the variance of the residuals from the fit, which would be 1.0 for a good fit. There are three outlier with residuals $\ge 3\,\sigma$ in the fit:   RR~Gem, an RRab star with $[\mathrm{Fe/H} = -0.29$, $\log P = -0.40$, $R = 3.6$; 
 TT~Lyn, an RRab star with  $[\mathrm{Fe/H} = -1.56$, $\log P = -0.22$, $R = 3.4$; and RU~Psc, an RRc star with 
  $[\mathrm{Fe/H} = -1.75$, $\log P = -0.41$, $R = 3.0$.  There is no indication from the photometry or the DR2 astrometric $\chi^2$ that the  parallaxes of these three stars are in error.  However, since in a sample of 143 stars, one does not expect any $\ge 3\,\sigma$ outliers, we also performed the fit with these  three stars removed.  The goodness of fit is improved when these three stars are removed. Removing the outliers does not lead to a significant change in the slopes of the fitted PLZ relation, and has a modest ($0.02\,$mag) effect on the zero-point.  
  
Regardless of exactly how we determine the [Fe/H] values of the stars, the fits are all quite similar, and agree with each other within their uncertainties.  Given our preference for the F1 metallicity scale,  
we determine the following PLZ relation
\begin{eqnarray}
M_{K_s}  & =&  (-2.8\pm 0.2)  (\log P +0.27)   \\ 
&+& (0.12\pm 0.02) ( [\mathrm{Fe/H}] + 1.3) - (0.41\pm 0.03) \nonumber
\label{eqnPLZ2}
\end{eqnarray}
with $\pi_{zp} = -0.042\pm 0.013\,$mas.

\begin{figure}
%aastex52 \epsscale{.80}
%\plotone{qq.pdf}
\plotone{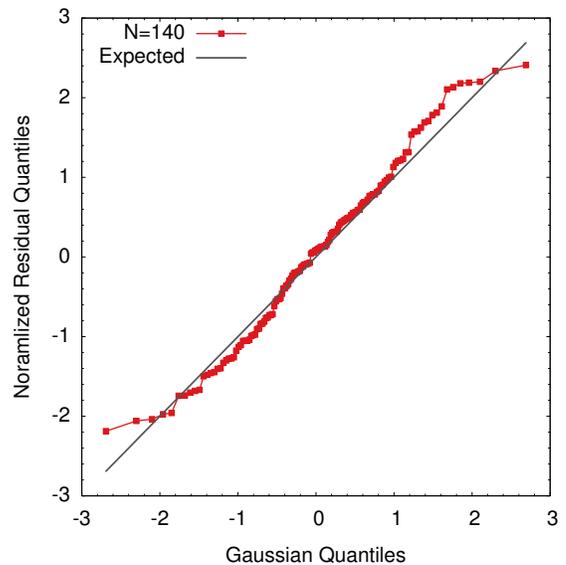}
\caption{A quantile-quantile plot of the normalized residuals for the fit to the stars in the F1 [Fe/H] scale, with outliers removed. A good fit will result in Gaussian residuals, with a standard deviation of $\sigma = 1$ and a mean $\mu = 0$, which is shown by the straight  line. 
    \label{figqq}}
\end{figure}

The quality of the fit may be visually  assessed using a  quantile-quantile plot (see Figure\ \ref{figqq}) to examine the normalized residuals.  It has been shown that the quantile-quantile plot is  superior to many other techniques (e.g.\  Anderson-Darling, Shapiro-Wilk or Cram\'{e}r-von Mises tests) in determining if a distribution is Gaussian \citep{loy}. In the quantile-quantile plot, the quantiles of the observed distribution are plotted as a function of the quantiles of the theoretical distribution. The slope of the quantile-quantile plot is the standard deviation  of the observed distribution, and the intercept is the mean of the observed distribution.  Figure \ref{figqq} shows the residuals of the F1 fit which does not include the three outliers (RR~Gem, TT~Lyn and RU~Psc).  The standard deviation of the residuals (slope in Fig.\ \ref{figqq}) is a bit larger than one, which suggests that there is some additional dispersion in the data which has not been captured by the fit.  

\begin{figure}
%aastex52 \epsscale{.80}
%\plotone{FitPLZ.pdf}
\plotone{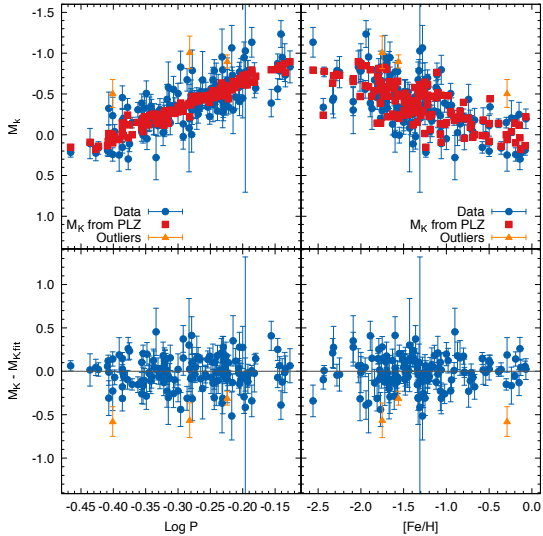}
\caption{The PLZ relation projected onto the PL (left) and PZ planes (right).  The upper panels show the $M_K$ values derived by assuming the distance is simply related to the parallax (Equation \ref{eqnMk}, including the DR2 zero-point error found by the fit) along with the $M_K$ values derived from the PLZ fit from Equation~\ref{eqnPLZ2}, and the lower panel shows the absolute magnitude residuals.    
    \label{fig_PLZ}}
\end{figure}

Figure  \ref{fig_PLZ} plots projections of the PLZ relation and residuals in the $\log P$-$M_{K_s}$ and [Fe/H]-$M_{K_s}$ planes. The three outliers are shown in orange, and were not included in the fit.  There are no obvious residual trends.

Our PLZ relation in $K_s$ may be compared to other determinations in the literature. Using data for 396 field RRL with DR2 parallaxes along with the single-epoch $K_s$ apparent magnitudes and reddening values from \citet{dambis13}, \cite{muraveva18} found 
\begin{eqnarray}
M_{K_s} &=& (-2.44\pm 0.35)  \log P  \\
&+& (0.18\pm 0.07)  [\mathrm{Fe/H}]  - (0.81\pm 0.18) \nonumber
\end{eqnarray}
and a DR2 global parallax error of $\pi_{zp} = -0.051\pm  0.01$ mas. At $ [\mathrm{Fe/H}]  = -1.3$ and $\log P = -0.27$, their relation implies $M_{K_s} = -0.40\pm 0.18\,$mag which agrees very well with the zero-point of our PLZ relation.  Our period slope is somewhat steeper (by  $0.89\,\sigma$)  than the \cite{muraveva18} period slope, while our [Fe/H] slope differs by $0.82\,\sigma$. 
The uncertainties given by \cite{muraveva18} are all larger than our uncertainties, except for the uncertainty in $\pi_{zp}$, which is slightly better constrained  in their analysis. \cite{muraveva18} performed a Bayesian analysis, and it is unclear if the differences between our results and \cite{muraveva18} are due to the different datasets employed, or the different analysis techniques. Our approach of ``centering'' the period and metallicity distributions by applying shifts of 0.27 and 1.3 respectively, minimizes the formal uncertainty in the fitted zero point, $\beta_1$.   It is worth noting that  our set of RRL is a subset, comprising most of the brighter, best-measured stars in the samples of \citet{muraveva18} and \citet{dambis13}, but using different apparent magnitudes, reddenings, and [Fe/H] values. 

Our [Fe/H] slope agrees with that found by \cite{braga18},  who determined slopes ranging from $0.062$ to $0.152$ depending on which set of [Fe/H] estimates they used in their fits.  \cite{braga18} found a period slope of $-2.3\pm 0.1$, which is $2.2\,\sigma$  shallower than our slope.  Our period slope is in agreement with  
 \cite{muraveva15}, who analyzed 70 RRL in the LMC and found a period slope of $-2.73\pm 0.25$, while their  [Fe/H] slope of $0.03\pm 0.07$ is $1.2\,\sigma$ shallower than our [Fe/H] slope. From theoretical models, \cite{marconi15} predict $M_K = -2.25\log P + 0.18 [Fe/H] - 0.82$ while \cite{catelan04} predict $M_K = -2.35\log P + 0.175\log Z - 0.60$.  In both cases, our period slope is steeper and our [Fe/H] slope is shallower than the theoretical predications. The differences are %of order 
 $\sim 2 \sigma$ for the period slope and $\sim 3 \sigma$ for the [Fe/H] slope.  It will be interesting to see if the slopes we find with the DR2 dataset are confirmed using the forthcoming \textit{Gaia} data releases; if they are confirmed, this would indicate a possible problem with the theoretical models.

%% Possible note about other tests we tried
 We explored subdividing our sample at $[\mathrm{Fe/H}] \approx -0.90\,$dex into disk and halo groups (see Sec.~\ref{ssec_samchar}) to test whether these kinematically distinct populations  -- which likely have different ages and chemical evolution histories -- have distinct luminosities.  However, the number of disk stars in our current sample was too small to obtain an acceptable fit.  Similar attempts to subdivide the halo sample into groups based on their period-shift behaviors, following the methods of \citet{leecarney99}, were also compromised by the small number of stars available which led to poor fits with large uncertainties and non-Gaussian residuals.  We were thus unable to test whether these field RRL, analogous to the Oosterhoff~I and II groups found in Galactic globular clusters, have distinct luminosities.  As the \textit{Gaia} mission continues and 
high-quality parallaxes become available for larger samples of stars, it will be interesting to see how the PLZ results for the disk, Oosterhoff~I and Oosterhoff~II samples develop.

\section{Summary}    \label{sec_concl}

%We note that \citet{dambis13} compiled a large list of 
%$\langle K_s \rangle$             **** this will be suggestions for future K-band photom
%$K_s$ values for 403 field RRL by phasing the single-epoch 2MASS photometry and applying template fitting.  We suspect these data, which include stars somewhat fainter than those in Table~\ref{tab_targets}, will become increasingly valuable for refining $M_K$(RR) as future \textit{Gaia} data releases provide parallaxes with even higher precisions.  [TO DO: add cautions about 15\% outliers and need to identify them; discuss flagged stars in our Table~\ref{tab_gaia}.]

We obtained over one thousand photometric measurements in the $K_s$ bandpass  for 74 field RRL  spread across the sky.  Our data are calibrated directly to the 2MASS photometric system using on-image comparison stars.  The RRL were selected to be relatively bright, with low interstellar reddenings, and to span a wide range of metallicity.  We favored stars with poor or no multi-epoch $K$-band photometry in the literature, and tended to avoid stars known to exhibit the Blazhko effect.  

We fit the phased observations for each star with a set of templates \citep{jcf96} to obtain the star's intensity-mean magnitude in $K_s$.  We checked the phase shift produced by the fit by comparing with the phasing of (near) contemporaneous optical photometry.  In cases where there were too few observations to constrain the fit (eight of  our 74 RRL stars), we used optical photometry to determine the phase and amplitude of the appropriate template and fitted to find the intensity-mean magnitude.  We provide a careful analysis of the resulting photometric uncertainties; the typical uncertainty in the intensity-mean magnitude is 0.029 mag for stars observed with SMARTS, and 0.013 mag for stars observed from MDM.

We compared our data with multi-epoch photometry in the literature after correcting it to the $K_s$ system.  We found excellent agreement between our data and the high quality light curves obtained for Baade-Wesselink analyses, and generally good agreement, though with larger scatter, was found with the more sparse light curves of \citet{fernley98} and the phased, single-epoch photometry of \citet{dambis13}.  There is some evidence that our photometry taken with SMARTS is systematically too bright by 0.02-0.03 mag, while that of \citet{dambis13} may be too faint by a similar amount; we encourage independent confirmation of this.  We found that $\sim$10\% of the stars from sources utilizing few observations \citep{fernley98,dambis13} are outliers by $\sim$0.1 mag, perhaps because the ephemerides used to phase these stars' photometry were outdated.  An additional 8\% of the RRL in \citet{dambis13}, generally ones fainter than those in our sample, are expected to suffer larger scatter because optical ephemerides were not available to allow their single-epoch 2MASS photometry to be phased and fit with a template. Together, these outliers may be a fundamental source of error in forthcoming studies of $M_K$(RR)
%RRL PLZ 
relations which use new data released by \textit{Gaia}.  We combined our photometry with these other sources using outlier rejection to provide a catalog of 146 RRL with reliable $K_s$ photometry.

We obtained estimates of the interstellar reddening toward each RRL from up to three independent sources and combined them to obtain a value for the interstellar  extinction  %absorption 
of each star.  These are in excellent agreement with the independent values of \citet{dambis13}.

We used the \textit{Gaia} DR2 parallaxes for these stars to calculate their absolute magnitudes, and analyzed them using the astrometric based luminosity prescription of \cite{ABL}, our implementation of which determines the global zero-point parallax error in DR2 for our sample.  We obtained a RRL PLZ relationship (see Equation~\ref{eqnPLZ2}) by including an intrinsic dispersion in $M_{K_s}$ of 0.04 mag which we estimated from the RRL in $\omega$ Centauri.  %We estimated the uncertainties in the PLZ coefficients using Monte Carlo simulations, [BC ADD?] %and found that our fitting procedure tended to underestimate the metallicity slope.  This probably results from the strong correlation between period and metallicity shown in Figure~\ref{fig_zper}, and may arise in other PLZ solutions that simultaneously fit these two properties.  After exploring the effects of fixing the metallicity coefficient at values found in the literature, we obtained our best estimate of the PLZ relation based on our data (see Equation~\ref{eqnPLZbest}). 
%This result is in good agreement with other PLZ relations in the literature that used \textit{Gaia} DR2 parallaxes as well as other approaches.  \textbf{ Three of our RRL were outliers in our fits (RR~Gem, TT~Lyn, and RU~Psc) for reasons we cannot identify.  All three are about 0.5 mag more luminous than typical stars of their period and metallicity, so they could be evolving across the instability strip at high luminosities from initial, helium core-burning loci as blue horizontal branch stars.  We note that one of these stars, TT~Lyn, has been associated with a stream of halo stars (Kinman et al. 2012) thought to be the relic of an ancient merger of a modest-sized galaxy with the Milky Way (Helmi et al. 1999). }
We estimated the uncertainties in the PLZ coefficients using Monte Carlo simulations,  and the results were in good agreement with the uncertainties reported by our implicit non-linear fitting routines, giving us confidence that the reported uncertainties in our PLZ relationship are realistic.  Our PLZ relation is in reasonable agreement with other PLZ relations in the literature that used DR2 parallaxes, as well as other approaches to determining the PLZ relation.  We identified three stars that were outliers from this fit: if they remain outliers when analyzed using parallaxes from future \textit{Gaia} data releases, then their photometry, [Fe/H] and reddening estimates should be reexamined.  

% \bibitem[Kimnan et al.(2012) {kinman12} Kinman, T. D., Cacciari, C., Bragaglia, A., Smart, R., and Spagna, A. 2012, \mnras, 422, 2116
% BHB and RRL in anticenter field, finds streams from Helmi99 and K07.

% \bibitem[Helmi et al.(1999) {helmi99} Helmi, A., White, S. D., de Zeeuw, P. T., and Zhao, H. 1999, Nature , 402, 53

%Finally, we searched for luminosity differences between different subsets of our RRL sample.  We divided the stars with halo kinematics into two groups corresponding to Oosterhoff~I and II RRL using their period-shift values.  Though there are currently too few stars in the analysis to allow reliable formal fits, we find evidence that the Osterhoff~II group may be more luminous, as expected if these stars have evolved into the instability from the blue horizontal branch.  Similarly, there are currently too few stars with thick disk kinematics to enable a separate PLZ fit for these stars.  We also explored fitting the data with a relation between period, luminosity and period-shift, and found a correlation of similar quality to our PLZ relation, though again the determination of the coefficients is complicated by the correlation between period and period-shift.  These $M_K$(RR) relations should continue to be investigated as the \textit{Gaia} mission continues and high-quality parallaxes become available for larger samples of RR~ Lyrae stars.

\acknowledgments
% We are grateful for the feedback from an anonymous referee, whose comments resulted in substantial improvements in this study.
We thank the anonymous referee for valuable comments that led to substantial improvements in this study.  
We are grateful to Bruce Carney, a member of our original \textit{SIM-PQ} team, for his valuable insights on star selection early in this work, and for his thoughtful comments on this manuscript.
We deeply appreciate the efforts of former SMARTS Queue Manager Rebeccah Winnick for %working tirelessly to 
scheduling our observations with timings that resulted in light curves with well-distributed phases.  We thank BGSU undergraduate Neil Beery for his patient work in generating finder charts,
%We thank former SMARTS Queue Manager Rebeccah Winnick ?? for working tirelessly to schedule our observations with timings that resulted in light curves with well-distributed phases, 
and Dr. Richard Pogge for his quick and helpful comments regarding photometric calibration of images taken with the SMARTS 1.3-m telescope and ANDICAM instrument.  
%We thank BGSU undergraduate Neil Beery for his work in generating finder charts.
%This work represents partial fulfillment of a Bachelor of Science degree in Physics at BGSU for CxB.  
This work was supported in part by award 1226828 from the Jet Propulsion Laboratory, in support of NASA's  \textit{Space Interferometry Mission}.
% Key Project %on Globular Clusters titled \textit{Anchoring the Population~II Distance Scale: Accurate Ages for Globular Clusters and Field Halo Stars}.  
%\textbf{ [TO DO: fill in grant info with correct acknowledgement]. }
This research has made use of the NASA/IPAC Infrared Science Archive, which is operated by the Jet Propulsion Laboratory, California Institute of Technology, under contract with the National Aeronautics and Space Administration.  This research has also made use of the VizieR catalogue access tool, CDS,
 Strasbourg, France (the original description of the VizieR service was
 published in \aap , 143, 23).  
The authors gratefully acknowledge the services of the AAVSO International Database (AID); some of the data acquired therefrom was made available by the British Astronomical Association Variable Star Section.

% Include Neil Beery as coauthor or in acknowledgements for help arranging finder charts.

%This research was made possible through the use of the AAVSO Photometric All-Sky Survey
%(APASS), funded by the Robert Martin Ayers Sciences Fund.  VSX, PROMPT. 
%


\vspace{5mm}
\facilities{CTIO:1.0m, McGraw-Hill, IRSA, 2MASS}


\software{IRAF \url{http://iraf.noao.edu}
%        Cloudy \citep{2013RMxAA..49..137F}, 
 %         SExtractor \citep{1996A&AS..117..393B}
          }



\appendix

In this Appendix, we provide additional discussion on the photometry and light curves of selected individual stars.

%Here we discuss the light curves and template fitting details for individual stars.

%WY~Ant: 

\textbf{ AA~Aqr:} Only three observations were obtained so the light curve fit was poorly constrained.  We determined a phase-shift from the maximum light of $I$-band observations obtained from BGSU less than one year before the MDM observations, and utilized template ``ab2'' with $\Delta K = 0.32$ mag as determined from Eqn.~7 of \citet{jcf96}, based on the observed amplitude $\Delta V = 1.14$ mag from \citet{munari14} and Eqn.~11 of \citet{jcf96}.  The resulting light curve fit appears reasonable (see Figure~\ref{fig_lc2}), with $\sigma_{fit} = 0.033$ mag, though the 2MASS magnitude for AA~Aqr is 0.01 mag fainter than the faintest magnitude of our fitted template, suggesting that a larger amplitude is possible.

\textbf{ BV~Aqr:} Both the original and VSX periods resulted in phased $I$-band light curves with more scatter than expected, so we used the $I$-band data to perform a period search, obtaining 0.36388 days.  This produced better light curves in both $I$ and $K_s$, though the latter still has significant scatter because a single, relatively faint comparison star was used to calibrate this variable.

\textbf{ S~Ara:} The $I$-band light curve shows evidence of the Blazhko effect, while the $K_s$ light curve does not show much scatter.

\textbf{ TW~Boo:}  Only two $K$-band observations were made using four very faint comparison stars.  We used $I$-band data taken two years prior to the MDM observations to determine a phase shift, and we used the $B$-band amplitude of 1.33 mag \citep{bookmeyer77} to set the $K$-band amplitude of template ``ab2'' to 0.31 mag, according to Eqn.~7 of \citet{jcf96}.  The resulting fit was adequate, but we treat the resulting $\langle K \rangle$ magnitude of this star with low confidence.

\textbf{ RV~Cap:} This star was observed using both SMARTS and MDM.  SMARTS: Our observed light curve shape is not well-matched by any of the templates; there is a sharp, early peak at maximum light and a rather flat or rounded minimum. We used the observed values for $K_{max}$ and $K_{min}$, but the observed and fitted intensity means were nearly identical.  MDM: Only two $K$-band observations were obtained.  We used BGSU $I$-band data to determine the phase shift, $\Delta \phi = -0.07$, and the $B$-amplitude of 1.71 \citep{bookmeyer77} to fix the ``ab4'' template with an amplitude of $\Delta K = 0.35$ mag.  The resulting one parameter fit seems reasonable, but the 2MASS $K_s$ magnitude is 0.11 mag brighter than the fitted template, and so we did not use the MDM observations further.  This star is known to exhibit the Blazhko effect \citep{smith95}, which has complicated establishing the phases and determining the time-averaged mean magnitude from the SMARTS data, and might account for the unusual light curve shape we obtained.

\textbf{ YZ~Cap:} This star was observed using both SMARTS and MDM.  SMARTS: We found this RRc variable to have a very small amplitude.  The 2MASS $K_s$ magnitude was slightly outside the range of our data, suggesting there may be a problem with the calibration which is based on two faint comparison stars.  MDM: The wider field of view allowed us to use more comparison stars and obtain a better light curve.  The rounded light curve and low amplitude resulted in three-parameter fits that were poorly constrained, so we determined the phase from $I$-band data taken at BGSU 1-2 years before the MDM observations.  The two-parameter fit solved for both the amplitude and zero-point, and appears trustworthy.  Ultimately, we combined the results from the two observatories using a weighted mean to obtain $\langle K_s \rangle = 10.355 \pm 0.011$ mag.

\textbf{ RX~Cet:} The data have a phase gap between $\phi = 0.76$ and 0.99, leaving the region around minimum light poorly constrained.  The initial three-parameter fit found a low amplitude of 0.18 mag, and the peak of the $I$-band light curve fell at $\phi = 0.04$.  A two-parameter fit where the phase was shifted by -0.04 still had a low amplitude of 0.20 mag.  In both cases, the 2MASS photometry was about 0.1 mag fainter than our observations and our fit.  The star's $B$-band amplitude of 1.13 mag \citep{bookmeyer77} suggests a $K$-amplitude of 0.29 mag according to Eqn.~7 of \citet{jcf96}, along with template``ab2.''  We performed a one-parameter fit using this template and amplitude along with $\Delta\phi = -0.04$, and obtained a good quality fit for which the 2MASS observation is within the range of the fitted template.  The intensity means of the three-, two- and one-parameter fits were $\langle K_s \rangle = 10.147$, 10.152, and 10.159 mag, respectively.   We adopt the results of the one-parameter fit.

%{bookmeyer77} Bookmeyer, B. B., Fitch, W. S., Lee, T. A., Wisniewski, W. Z., \& Johnson, H. L. 1977, Rev. Mex. Astr. Ap., 2,  235

%RY~Col: No comp stars -- SKIP

\textbf{ DM~Cyg:} While the template fit to the variable star light curve was excellent and the non-variable check star was in good agreement with its 2MASS $K_s$ magnitude, the 2MASS magnitude for the variable was 0.07 mag brighter than the magnitude range of our fitted template.  This may result from the known Blazhko behavior of this star \citep{smith95}.

\textbf{ XZ~Dra:} Both the original and VSX periods phased our $I$-band photometry, taken at BGSU 4-5 years prior to the MDM data, so that $I_{max}$ occurred at $\phi = 0.14$, not zero.  We used VSTAR to search for a better period, 0.47648 days, which, when used in conjunction with the $K$-band data, resulted in a high quality three-parameter fit with $I_{max}$ at zero phase.    We note, however, that XZ~Dra is one of the two stars that failed the cycle-count test described in Sec.~\ref{ssec_mdmfit} -- we estimated the discrepancy in the formal cycle-count to be 0.24 cycles.  This, together with the fact that XZ~Dra is known to exhibit the Blazhko effect \citep{smith95}, lead us to suggest further $K_s$ photometry be obtained for this star. 

\textbf{ BX~Dra:} This star was originally classified as an ab-type RR Lyrae star in the sources we used for our initial target list, but was since reclassified as a W~Ursae Majoris contact binary.  Our data on this star were fit nicely with a template derived from a W~UMa star's light curve, thus confirming the new classification.

\textbf{ VX~Her:} Partial phase coverage by the $K$-band data required a one-parameter fit using $\Delta\phi = -0.09$ derived from $V$-band data taken at BGSU the same years as the MDM data. The star's $B$-band amplitude, 1.45 mag \citep{bookmeyer77}, indicates a $K$-band amplitude of 0.34 mag according to Eqn.~7 of \citet{jcf96}, along with template``ab4.''  The resulting template fit the data well, though the 2MASS observation was 0.02 mag brighter than maxiumum of the fitted template, suggesting template ``ab5'' might be appropriate.

%SS~For: The light curve was noisy because only one very faint comparison star was usable.

\textbf{ SX~For:} The light curve was noisy because only one very faint comparison star was usable.  The 2MASS $K_s$ magnitude value was 0.16 mag fainter than the range of our fitted template, calling the calibration into question.

%The light curves for the following stars were noisy because only one very faint comparison star was usable: SS~For, ...?, VX~Scl.

\textbf{ VV~Peg:} The $K$-band data covered only a small range in phase, leaving the level of the minimum unconstrained.  Based on $V$-band data taken at BGSU two months later, we determined a phase shift  $\Delta\phi = -0.14$ should be applied.  The $B$-band amplitude of the star is 1.45 mag \citep{bookmeyer77} indicating a $K$-band amplitude of 0.33 mag according to Eqn.~7 of \citet{jcf96}, along with template ``ab3.''  We performed a one-parameter fit using these values and obtained a good result for which the 2MASS observation was within the range of the fitted template.

\textbf{ W~Tuc:} This star had one very faint comparison star which was only useable when the seeing was good, so only nine images resulted in photometry.  The small number of observations did not constrain the phase well, so we set the phase so the SMARTS $I$-band data peaked at $\phi = 0$.  The large scatter resulted in selection of an unrealistically-small amplitude, so we set the amplitude to $\Delta K = 0.35$ and utilized the ``ab5'' template based on the results of \citet{cacciari92}, and did a one-parameter fit for the zero-point magnitude.  The large scatter in the resulting fit leaves us dubious of the usefulness of this data, but we include it for completeness.

\textbf{ RV~UMa:} Two $K$-band observations were obtained.  A phase shift was determined from $V$-band data taken at BGSU 3-7 years before the MDM observations.  The amplitude in  $B$ of 1.47 mag from \citet{bookmeyer77} indicates a $K$-band amplitude of 0.33 mag with template ``ab3.''  The resulting one-parameter fit was extremely good.     However, RV~UMa is the second star that failed the cycle-count test described in Sec.~\ref{ssec_mdmfit}, having a formal cycle-count discrepancy of 0.28 cycles.  As for XZ~Dra, we suggest further $K_s$ photometry be obtained for this star.

\end{document}